\pdfoutput=1
\RequirePackage{ifpdf}
\ifpdf 
\documentclass[pdftex]{sigma}
\else
\documentclass{sigma}
\fi

\numberwithin{equation}{section}

\newtheorem{Theorem}{Theorem}[section]
\newtheorem*{Theorem*}{Theorem}

 { \theoremstyle{definition}

\newtheorem{Remark}[Theorem]{Remark} }

\newcommand{\pa}{\partial}
\newcommand{\nn}{\nonumber}

\begin{document}
\allowdisplaybreaks

\newcommand{\arXivNumber}{2210.12771}

\renewcommand{\PaperNumber}{015}

\FirstPageHeading

\ShortArticleName{Stationary Flows Revisited}

\ArticleName{Stationary Flows Revisited}

\Author{Allan P.~FORDY~$^{\rm a}$ and Qing HUANG~$^{\rm b}$}

\AuthorNameForHeading{A.P.~Fordy and Q.~Huang}

\Address{$^{\rm a)}$~School of Mathematics, University of Leeds, Leeds LS2 9JT, UK}
\EmailD{\href{mailto:a.p.fordy@leeds.ac.uk}{a.p.fordy@leeds.ac.uk}}
\URLaddressD{\url{https://eps.leeds.ac.uk/maths/staff/4024/professor-allan-p-fordy}}

\Address{$^{\rm b)}$~School of Mathematics, Center for Nonlinear Studies, Northwest University,\\
\hphantom{$^{\rm b)}$}~Xi'an 710069, P.R.~China}
\EmailD{\href{mailto:hqing@nwu.edu.cn}{hqing@nwu.edu.cn}}

\ArticleDates{Received October 28, 2022, in final form March 08, 2023; Published online March 29, 2023}

\Abstract{In this paper we revisit the subject of stationary flows of Lax hierarchies of a coupled KdV class. We explain the main ideas in the standard KdV case and then consider the dispersive water waves (DWW) case, with respectively 2 and 3 Hamiltonian representations. Each Hamiltonian representation gives us a different form of stationary flow. Comparing these, we construct Poisson maps, which, being non-canonical, give rise to bi-Hamiltonian representations of the stationary flows. An alternative approach is to use the Miura maps, which we do in the case of the DWW hierarchy, which has two ``modifications''. This structure gives us 3 sequences of Poisson related stationary flows. We use the Poisson maps to build a tri-Hamiltonian representation of each of the three stationary hierarchies. One of the Hamiltonian representations allows a multi-component squared eigenfunction expansion, which gives $N$ degrees of freedom Hamiltonians, with first integrals. A Lax representation for each of the stationary flows is derived from the coupled KdV matrices. In the case of 3 degrees of freedom, we give a generalisation of our Lax matrices and Hamiltonian functions, which allows a connection with the rational Calogero--Moser (CM) system. This gives a coupling of the CM system with other potentials, along with a Lax representation. We present the particular case of coupling one of the integrable H\'enon--Heiles systems to~CM.}


\Classification{35Q53; 37K10; 70H06}

\section{Introduction}
It has long been known \cite{76-5} that the stationary flows of nonlinear evolution equations of ``KdV-type'' are themselves (finite-dimensional) completely integrable Hamiltonian systems. Specifically, \cite{76-5} considered the KdV hierarchy, with its first Hamiltonian structure:
\begin{gather}\label{kdv-dxdelh}
u_{t_n} = \pa_x \delta_u H_{n+1},
\end{gather}
where $H_n$ are the KdV Hamiltonian densities. Setting $u_{t_n}=0$ leads to an ODE in (generalised) Lagrangian form, with ${\mathcal L}_{n+1}=H_{n+1}-\alpha u$. The (generalised) Legendre transformation then leads to canonical variable, in which the stationary flow takes Hamiltonian form. The lower flows~$u_{t_m}$, $m=1,\dots ,n-1$, then restrict to this finite-dimensional manifold, forming a system of commuting Hamiltonian flows, corresponding to Hamiltonian functions which are derived from the fluxes of the lower KdV densities.

In \cite{f87-3}, it was shown that each of these stationary flows is {\em bi-Hamiltonian} on an extended space, with additional {\em dynamical variable} $\alpha$. This was derived by using the Miura map, which also gave a bi-Hamiltonian formulation of the {\em stationary} MKdV equation. This should be compared with the full MKdV equation, with only {\em one local} Hamiltonian structure.

In \cite{f91-1}, the 3 known integrable cases of the H\'enon--Heiles equation were shown to be related to the {\em stationary reductions} of the 3 known fifth order (single component) nonlinear evolution equations with a Lax representation. One of these is in the KdV hierarchy, but not in the canonical coordinates derived from (\ref{kdv-dxdelh}). In this calculation the fourth order stationary flow (for one coordinate $q_1$) is derived as a consequence of the coupled {\em second order} system, by differentiation and elimination, so the relation to canonical variables is not addressed. In \cite{f95-3}, these canonical variables are {\em directly constructed} by using the {\em second} Hamiltonian structure of the KdV hierarchy. Indeed, in \cite{f95-3} a general approach, using squared eigenfunction coordinates, was introduced, relating several well-known completely integrable,
finite-dimensional Hamiltonian systems to the stationary flows of various well-known Lax hierarchies.\looseness=1

In this paper we revisit this subject, bringing together the ideas of \cite{f87-3} and \cite{f95-3}, enabling us to build bi- and tri-Hamiltonian representations of a number of interesting systems. The approach also allows us to build Lax representations for these. These systems are certainly {\em completely integrable}, but some are {\em superintegrable} \cite{14-2,90-22,13-2}.

Since many of our formulae follow from those of {\em coupled KdV hierarchies}, associated with ``energy-dependent'' Schr\"odinger operators \cite{f87-5,f89-2}, we give a brief overview of these in Section~\ref{Lax-background}.\looseness=1

Our main explanation of {\em stationary reductions} is given in Section \ref{sec:KdV-stat}, in the context of the KdV hierarchy. The two Hamiltonian representations of each PDE flow give us two stationary manifolds with different coordinates. This can be compared with the approach of \cite{f87-3}, which used the Miura map and the ``modified'' equation. The second of the Hamiltonian representations in this paper has a multi-component generalisation which gives some superintegrable systems, such as the {\em Garnier system} (\ref{L1-t1-phi}) and a generalised {\em H\'enon--Heiles system} (\ref{L2-ht}) (both with $N$-components). Choosing $N$ so that the stationary manifolds of the two representations have the {\em same} dimension, allows us to build Poisson maps between these manifolds, giving us a~bi-Hamiltonian representation of the stationary flows (see Sections~\ref{sec:biHam-KdV-t1} and~\ref{sec:biHam-KdV-t2}). The Lax representations of these stationary flows are discussed in Section~\ref{sec:sKdV-Lax}.\looseness=1

We discuss the stationary flows of the DWW hierarchy in Section \ref{sec:DWW-stat}. Following the ideas of Section \ref{sec:KdV-stat}, we find $N$-component, superintegrable systems (\ref{DWW-B1-t1-ht}) and (\ref{DWW-B1-t2-ht}), with (respectively) sextic and quartic potentials. Again, choosing $N$ so that the stationary manifolds of the two representations have the {\em same} dimension, allows us to build Poisson maps, giving us bi-Hamiltonian representations (see Sections \ref{sec:DWW-N=1biHam} and \ref{sec:DWW-N=2biHam}). Lax representations are presented in Section \ref{sec:sDWW-Lax}.\looseness=1

In Section \ref{sec:DWW-triH}, we use the Miura maps of Section \ref{sec:DWW-Miura} to construct a {\em tri-Hamiltonian} formulation of the {\em stationary flows} of the DWW hierarchy and the two {\em modified hierarchies}, depicted in Figure \ref{B2B1B0-fig}. As this figure indicates, there are six {\em local} Hamiltonian operators (for the PDEs), arranged in a triangular array. In the stationary coordinates, we build a square array of {\em nine} Poisson matrices, so each of the {\em stationary} hierarchies is {\em tri-Hamiltonian}.

Some of the Lax representations of Sections \ref{sec:sKdV-Lax} and \ref{sec:sDWW-Lax} can be generalised to incorporate an arbitrary function, which enable us to couple some of our potentials to the {\em rational Calogero--Moser system} \cite{71-3,75-3}. This is discussed in Section \ref{sec:generalisations}, where we specifically generalise the Garnier system (\ref{L1-t1-phi}), the generalised {\em H\'enon--Heiles system} (\ref{L2-ht}) and the system with Hamiltonian~(\ref{DWW-B1-t2-ht}). Using the approach of \cite{f22-1}, these can all be canonically transformed to the {\em rational Calogero--Moser} Hamiltonian, with additional potential terms. These are all completely integrable and have a Lax representation. To illustrate this, we explicitly present the H\'enon--Heiles case.

\section{Lax representation of coupled KdV equations}\label{Lax-background}

In \cite{f87-5,f89-2} a detailed analysis was given of coupled KdV equations, associated with ``energy dependent'' Schr\"odinger operators. Here we give a very brief review to present a few facts which we use in this paper.

The ``energy dependent'' Schr\"odinger equation is
\begin{subequations}
\begin{gather}\label{e-Lax}
L \psi = \big(\pa_x^2+u\big) \psi = 0 , \qquad \text{where}\quad u = \sum_{i=0}^{M-1} u_i \lambda^i -\lambda^M,
\end{gather}
such that $u_i$ are functions of $x$ and some ``time'' parameters $t_n$. Suppose for one of these (denoted~``$t$''), we have
\begin{gather}\label{psit}
\psi_t = A \psi = (2 P \pa_x - P_x )\psi.
\end{gather}
From (\ref{e-Lax}) and (\ref{psit}) we can obtain two formulae for $\psi_{xxt}$, which, when equated, lead to
\begin{gather}\label{compat2}
u_t=\big(\pa_x^3+4 u \pa_x+2 u_x\big) P.
\end{gather}
The coupled KdV hierarchy is given by the {\em polynomial expansion}
\begin{gather}\label{Pol-P}
P^{(n)} = \sum_{k=0}^n \lambda^{n-k} P_k, \qquad\text{with}\quad P_0=1.
\end{gather}
Substituting this into (\ref{compat2}) and equating coefficients of powers of $\lambda$ gives us a recursive formula for $P_k$ and then the formulae for $u_{it_n}$, which can be written in matrix form to give the Hamiltonian formulation. In this paper we only consider the KdV ($M=1$) and DWW ($M=2$) cases, so just describe the Hamiltonian formulations for these specific cases.
\end{subequations}

This differential operator Lax pair is rewritten in ``zero-curvature'' form
\begin{subequations}
\begin{gather}\label{zero-c}
\begin{pmatrix}
 \psi_1 \\
 \psi_2
 \end{pmatrix}_x =
 \begin{pmatrix}
 0 & 1 \\
 -u & 0 \\
 \end{pmatrix}
 \begin{pmatrix}
 \psi_1 \\
 \psi_2
 \end{pmatrix}=U\Psi,\qquad
 \begin{pmatrix}
 \psi_1 \\
 \psi_2
 \end{pmatrix}_{t}= \begin{pmatrix}
 - P_x & 2 P \\
 -2 u P- P_{xx} & P_x \\
 \end{pmatrix}
 \begin{pmatrix}
 \psi_1 \\
 \psi_2
 \end{pmatrix}=V \Psi ,
\end{gather}
with integrability condition
\begin{gather}\label{Ut}
U_{t}-V_x+[U,V]=0,
\end{gather}
which again leads to (\ref{compat2}). For a specific hierarchy, we must substitute the correct form of $u$. For the specific $t_n$ flow, we must also use~$P^{(n)}$ of~(\ref{Pol-P}) in $V$, leading to
\begin{gather}\label{Utn}
U_{t_n}-V^{(n)}_x+\big[U,V^{(n)}\big]=0.
\end{gather}
We then have a direct construction of a Lax pair for the {\em stationary flow} $U_{t_n}=0$, by rewriting~$V^{(n)}$ in terms of the coordinates being used on the stationary manifold. This will be explained in more detail in Section~\ref{sec:KdV-stat}.
\end{subequations}

\subsection{The KdV hierarchy and conservation laws}

The KdV hierarchy corresponds to choosing $M=1$ in (\ref{e-Lax}), so $u=u_0-\lambda$. The coefficients $P_k$ in (\ref{Pol-P}) are constructed recursively, the first few being
\begin{gather*}
P_0 = 1,\qquad P_1 = \frac{1}{2} u_0,\qquad P_2 = \frac{1}{8} \big(u_{0xx}+3u_0^2\big).
\end{gather*}
The flows are given by (\ref{compat2}), with $P=P^{(n)}$ and can be written in bi-Hamiltonian form
\begin{subequations}
\begin{gather}\label{KdV-utn}
u_{0t_n} = B_1 \delta_{u_0} H_n = B_0 \delta_{u_0} H_{n+1}, \qquad n\geq 0,
\end{gather}
with
\begin{gather}\label{KdV-B1B0}
B_1 = \pa_x^3+4 u_0 \pa_x+2 u_{0x},\qquad B_0 = 4 \pa_x \qquad\text{and}\qquad P_n = \delta_{u_0} H_n,
\end{gather}
where $\delta_{u_0}$ denotes the variational derivative with respect to $u_0$.

In particular,
\begin{gather}
 H_0 = u_0,\qquad H_1 = \frac{1}{4} u_0^2,\qquad H_2 = \frac{1}{8} \left(u_0^3-\frac{1}{2} u_{0x}^2\right), \nonumber\\
 H_3 = \frac{1}{64} \big(5 u_0^4-10 u_0 u_{0x}^2+u_{0xx}^2\big).\label{KdV-Hm}
\end{gather}
Corresponding to these densities, we have an array of {\em fluxes} $F_{nm}$, in local conservation laws, given by the $t_n$ evolution of $H_m$:
\begin{gather}\label{KdV-Hmtn}
\pa_{t_n} H_m = \pa_x F_{nm}.
\end{gather}
In particular, we have
\begin{gather}
F_{10} = \frac{1}{2} \big(u_{0xx}+3 u_0^2 \big),\qquad F_{11} = \frac{1}{8} \big(2 u_0 u_{0xx} -u_{0x}^2+4 u_0^3 \big),\nn \\
 F_{20} = \frac{1}{8} \big(10 u_0^3 +5 u_{0x}^2 +10 u_0 u_{0xx} + u_{0xxxx} \big), \nn\\
F_{21} = \frac{1}{32} \big(15 u_0^4 +20 u_0^2u_{0xx}+u_{0xx}^2-2 u_{0x}u_{0xxx}+2 u_0 u_{0xxxx} \big). \label{F12m}
\end{gather}
\end{subequations}

\subsection{The DWW hierarchy and conservation laws}

The DWW hierarchy corresponds to choosing $M=2$ in (\ref{e-Lax}), so $u=u_0+u_1 \lambda -\lambda^2$. The coefficients $P_k$ in (\ref{Pol-P}) are constructed recursively, the first few being
\begin{gather*}
P_0 = 1,\qquad P_1=\frac{1}{2} u_1,\qquad P_2=\frac{1}{8} \big(4 u_0+3 u_1^2 \big),\qquad P_3= \frac{1}{16} \big(12 u_0u_1+5 u_1^3+2 u_{1xx} \big).
\end{gather*}
The general recursion, defined by (\ref{compat2}) is
\begin{subequations}
\begin{gather}\label{DWW-JP=0}
J_0P_{m-2}+J_1P_{m-1}+J_2P_m=0,
\end{gather}
with
 \begin{gather*} J_0=\pa_x^3+4 u_0 \pa_x+2 u_{0x},\qquad J_1= 4 u_1 \pa_x+2u_{1x},\qquad J_2= -4\pa_x.
\end{gather*}
For $P=P^{(n)}$, we have
\begin{gather}\label{DWW-utn}
u_{0t_n}+\lambda u_{1t_n} = J_0P_n+\lambda (J_0P_{n-1}+J_1P_n),
\end{gather}
which can be written in vector form
\begin{gather}\label{DWW-ut=Bgrad}
 \begin{pmatrix}
 u_0 \\
 u_1 \\
 \end{pmatrix}_{t_n} = \begin{pmatrix}
 0 & J_0 \\
 J_0 & J_1 \\
 \end{pmatrix}
 \begin{pmatrix}
 P_{n-1} \\
 P_n \\
 \end{pmatrix} =
 \begin{pmatrix}
 J_0 & 0 \\
 0 & -J_2 \\
 \end{pmatrix}
 \begin{pmatrix}
 P_{n} \\
 P_{n+1} \\
 \end{pmatrix}
 =
 \begin{pmatrix}
 -J_1 & -J_2 \\
 -J_2 & 0 \\
 \end{pmatrix}
 \begin{pmatrix}
 P_{n+1} \\
 P_{n+2} \\
 \end{pmatrix},
\end{gather}
\end{subequations}
with the first being given directly by (\ref{DWW-utn}). The general recursion (\ref{DWW-JP=0}) then gives the second two representations. These matrix operators are the 3 compatible Hamiltonian operators of the DWW hierarchy, labelled respectively as $B_2$, $B_1$ and $B_0$. Defining
\begin{subequations}
\begin{gather}\label{DWW-delHn}
 \begin{pmatrix}
 P_{n-1} \\
 P_n \\
 \end{pmatrix}
 = \delta_u H_n = \begin{pmatrix}
 \delta_{u_0} H_n \\
 \delta_{u_1} H_n \\
 \end{pmatrix},
\end{gather}
we have
\begin{gather}\label{DWW-triHam}
 {\bf u}_{t_n} = B_2 \delta_u H_n = B_1 \delta_u H_{n+1} = B_0 \delta_u H_{n+2}, \qquad\text{for} \quad n\geq 0,
\end{gather}
where ${\bf u} = (u_0,u_1)^{\rm T}$. This is the construction given in \cite{f87-5,f89-2}. An $r$-matrix formulation was given in \cite{88-1}.

The above construction gives an infinite sequence of Hamiltonian functions, $H_n$, the first few of which are
\begin{gather}
H_0= u_1,\qquad H_1 = u_0+\frac{1}{4} u_1^2,\qquad H_2 = \frac{1}{2} u_1\left(u_0+\frac{1}{4} u_1^2\right), \nonumber\\
H_3 = \frac{1}{4}\left(u_0^2+\frac{3}{2} u_0u_1^2 +\frac{5}{16} u_1^4-\frac{1}{4}u_{1x}^2\right).\label{DWW-Hn}
\end{gather}
We have
\begin{gather}\label{DWW-BiCas}
B_0\delta_u H_0=B_0\delta_u H_1=B_1\delta_u H_0=0,
\end{gather}
meaning that $B_0$ has {\em two local Casimir functions} $H_0$, $H_1$, whilst $B_1$ has {\em one local Casimir function} $H_0$.

We define {\em fluxes} $F_{nm}$ by the same formula (\ref{KdV-Hmtn}), giving
\begin{gather}
 F_{10} = 2 u_{0}+\frac{3}{2} u_1^2,\qquad F_{11} = \frac{1}{2} \big(4 u_0 u_{1} + u_1^3+u_{1xx}\big), \nonumber\\
 F_{12} = \frac{1}{32} \big(16 u_0^2+40 u_0 u_1^2+9 u_1^4-4 u_{1x}^2+8 u_1 u_{1xx}\big), \qquad F_{20} = 3 u_0 u_{1} + \frac{5}{4} u_1^3+\frac{1}{2}u_{1xx},\nonumber\\
 F_{21} = \frac{1}{32} \big(48 u_0^2+72 u_0 u_1^2+15 u_1^4+20 u_{1x}^2+16 u_{0xx}+32 u_1 u_{1xx}\big), \nonumber\\
 F_{22} = \frac{1}{32} \big(48 u_0^2 u_1+9 u_1^5-8 u_{0x} u_{1x}+8 u_1 u_{0xx}+18 u_1^2 u_{1xx}+8 u_0 \big(6 u_1^3+u_{1xx}\big)\big).\label{DWW-F12m}
\end{gather}

\end{subequations}

\subsubsection{Miura maps}\label{sec:DWW-Miura}

In \cite{f89-2}, Miura maps were presented for the entire class of systems described by the Lax operator~(\ref{e-Lax}). In the DWW case, there are 3 sets of variables $(u_0,u_1)$, $(w_0,w_1)$ and $(v_0,v_1)$, related~by\looseness=-1
\begin{gather}
u_0 = -w_{0x}-w_0^2,\qquad u_1=w_1 \qquad\text{and}\qquad w_0=v_0,\qquad w_1 = -v_{1x}-2 v_0v_1 \nonumber\\ \qquad{} \Rightarrow \quad
 u_0=-v_{0x}-v_0^2, \qquad
 u_1=-v_{1x}-2 v_0 v_1.
 \label{mDWW}
\end{gather}
In the $u$-space we have the 3 {\em local} Hamiltonian operators $B_i^u \equiv B_i$, $i=0,1,2$ given by (\ref{DWW-ut=Bgrad}), with 2 {\em local} operators $B_2^w$, $B_1^w$ in the $w$-space and just 1 {\em local} operator $B_2^v$ in the $v$-space. These are depicted in Figure \ref{B2B1B0-fig} and related by
\begin{gather*}
B_k^u = \frac{Du}{Dw} B_k^w \left(\frac{Du}{Dw}\right)^\dagger ,\qquad\text{for}\quad k=1,2, \qquad\text{and}\qquad B_2^w = \frac{Dw}{Dv} B_2^v \left(\frac{Dw}{Dv}\right)^\dagger ,
\end{gather*}
where $\frac{Du}{Dw}$ and $\frac{Dw}{Dv}$ are the Jacobians of the maps (\ref{mDWW}), with
\begin{gather*}
B_2^w =\begin{pmatrix}
 0 & -\pa_x (\pa_x-2 w_0) \\
 (\pa_x+2 w_0)\pa_x & 4 w_1 \pa_x+2 w_{1x} \\
 \end{pmatrix},
\\
 B_2^v =\begin{pmatrix}
 0 & -\pa_x \\
 -\pa_x & 0 \\
 \end{pmatrix},
\qquad B_1^w =\begin{pmatrix}
 -\pa_x & 0 \\
 0 & 4\pa_x \\
 \end{pmatrix}.
\end{gather*}

\begin{figure}[t]\centering

\unitlength=0.5mm
\begin{picture}(80,40)
\put(0,40){\makebox(0,0){$B_2^u$}}
\put(0,20){\makebox(0,0){$B_1^u$}}
\put(0,0){\makebox(0,0){$B_0^u$}}
\put(40,40){\makebox(0,0){$B_2^w$}}
\put(40,20){\makebox(0,0){$B_1^w$}}
\put(80,40){\makebox(0,0){$B_2^v$}}
\put(30,40){\vector(-1,0){20}}
\put(30,20){\vector(-1,0){20}}
\put(70,40){\vector(-1,0){20}}
\end{picture}
\caption{Hamiltonian operators in the 3 spaces.} \label{B2B1B0-fig}
\end{figure}

We have built a sequence of Hamiltonians $H_k^u\equiv H_k^u[u_0,u_1]$, as functions of $u_0$, $u_1$ and their $x$-derivatives.
The Miura maps then define $H_k^w$ and $H_k^v$, by substituting the formulae (\ref{mDWW}) into~$H_k^u$.
The flow ${\bf u}_{t_n}$, defined by (\ref{DWW-triHam}), gives rise to the flows:
\begin{gather}\label{DWW-wtn-vtn}
 {\bf w}_{t_n} =B_2^w \delta_w H_n^w = B_1^w \delta_w H_{n+1}^w \qquad\text{and}\qquad {\bf v}_{t_n} =B_2^v \delta_v H_n^v,
\end{gather}
where ${\bf w}=(w_0,w_1)^{\rm T}$ and ${\bf v}=(v_0,v_1)^{\rm T}$.

\section{KdV stationary flows}\label{sec:KdV-stat}

In this section we discuss the two different canonical representations of the stationary flows, related to the two Hamiltonian representations of the KdV hierarchy. This gives us a new way of constructing a bi-Hamiltonian representation of the stationary flows. We also discuss the Lax formulation of the derived systems.

A {\em stationary} flow (for $t_n$) means that we reduce to a {\em finite-dimensional} space with $u_{0t_n}=0$. The ``time'' variable for this system is $x$, which is the variable which appears in the Lagrangians, given below.

This gives an ODE, defined by one of the two representations given in (\ref{KdV-utn}):
\begin{enumerate}\itemsep=0pt
 \item Using $B_0$ with the Bogoyavlensky--Novikov coordinates \cite{76-5}, we build the first Lagrangian:%
\begin{gather}\label{KdV-Lagn+1}
B_0 \delta_{u_0} H_{n+1}=0 \quad\Rightarrow\quad \delta_{u_0} (H_{n+1}-\alpha u_0)=0 \quad\Rightarrow\quad {\mathcal L}_{n+1} = H_{n+1}-\alpha u_0.
\end{gather}
We then use the (generalised) Legendre transformation to find canonical coordinates $(q_i,p_i)$ and the Hamiltonian function.
 \item Using $B_1$ and the squared eigenfunction representation (following \cite{f95-3}):
 \begin{subequations}
 \begin{gather}\label{KdV-dH=phi2}
\delta_{u_0} H_n = a \varphi^2,
\end{gather}
and since $B_1$ is {\em skew symmetric}, we can show
\begin{gather}\label{KdV-phixx}
\varphi^2 \big(B_1 \varphi^2\big) = \left(2 \varphi^3 (\varphi_{xx} +u_0 \varphi) \right)_x = 0 \quad\Rightarrow\quad 2 \varphi^3 (\varphi_{xx} +u_0 \varphi) = 4 \beta.
\end{gather}
Equations (\ref{KdV-dH=phi2}) and (\ref{KdV-phixx}) can then be written as variational derivatives of a single Lagrangian function
\begin{gather}\label{KdV-deltaLn}
{\mathcal L}_n = \frac{1}{2a}H_n - \frac{1}{2} u_0 \varphi^2+\frac{1}{2} \varphi_x^2-\frac{\beta}{\varphi^2}.
\end{gather}
\end{subequations}
Again, the (generalised) Legendre transformation gives canonical coordinates $(Q_i,P_i)$ and the Hamiltonian function.
\end{enumerate}

\begin{Remark}
Clearly, the algorithm for constructing the $H_n$ of (\ref{KdV-Hm}) leads to awkward looking overall factors. In the above formulae for ${\mathcal L}_{n}$ and ${\mathcal L}_{n+1}$, we always choose convenient multiples of $H_n$ and $H_{n+1}$.
\end{Remark}

\begin{Remark}
In equation (\ref{KdV-dH=phi2}), the parameter $a$ is arbitrary, but we can always rescale $\varphi$, so choose~${a=\frac{1}{2}}$.
\end{Remark}

Each of the above representations gives an $n$-degrees of freedom {\em canonical Hamiltonian} system for the stationary flow $u_{0t_n}=0$. We can extend the spaces to $2n+1$ dimensions, by adding the arbitrary constants $\alpha$ and $\beta$ as {\em dynamical variables}, which are just Casimir functions of the respective extended canonical brackets.

{\bf Poisson map.}
We can then write $q_i$, $p_i$, $\alpha$ in terms of $u_0(x)$ and its derivatives, which in turn can be written in terms of $Q_i$, $P_i$, $\beta$ and vice versa, thus giving us a {\em Poisson map} between the two systems. We emphasise that this is \emph{not} a {\em canonical transformation}, so generates a second Poisson bracket for each of the systems. We give a more detailed description of the procedure for the simplest case of the $t_1$ flow.

{\bf First integrals.}
Setting $u_{0t_n}=0$ in (\ref{KdV-Hmtn}) means that the fluxes $F_{nm}$ are {\em first integrals}. On this (extended) stationary manifold, the $n+1$ functions $\{F_{nk}\}_{k=0}^n$ are independent.

{\bf Multi-component squared eigenfunctions.}
Since $B_1$ is a linear operator, we can extend the squared eigenfunction representation (\ref{KdV-dH=phi2}) to include multiple eigenfunctions, with
\begin{subequations}
\begin{gather}\label{KdV-mult-phi}
\delta_{u_0} H_n = \frac{1}{2} \sum_i \varphi_i^2.
\end{gather}

{\it The first option} is to repeat the calculation (\ref{KdV-dH=phi2}) for each $\varphi_i$, to obtain $2 \varphi_i^3 (\varphi_{ixx} +u_0 \varphi_i) = 4 \beta_i$, giving
\begin{gather}\label{KdV-deltaLn-phii-1}
{\mathcal L}_n = H_n + \sum_i \left(\frac{1}{2} \varphi_{ix}^2 - \frac{1}{2} u_0 \varphi_i^2-\frac{\beta_i}{\varphi_i^2}\right) .
\end{gather}

{\it For the second option} we first define $\varphi^2 = \sum_i \varphi_i^2$. We then consider
\begin{gather}\label{KdV-f2-B1-f2}
\varphi^2 B_1 \varphi^2 = 0 \quad\Rightarrow\quad \varphi^2 \sum_i \varphi_i (\varphi_{ixx}+u_0 \varphi_i) + \sum_{i<j} (\varphi_i \varphi_{jx}-\varphi_j \varphi_{ix})^2 = \text{const}.
\end{gather}
If we now set
\begin{gather}\label{KdV-Lf-kf/f4}
\varphi_{ixx}+u_0 \varphi_i = \frac{2\beta\varphi_i}{\varphi^4}, \qquad\text{then}\quad \varphi^2 \sum_i \varphi_i (\varphi_{ixx}+u_0 \varphi_i) = 2 \beta.
\end{gather}
Furthermore, $(\varphi_i \varphi_{jx}-\varphi_j \varphi_{ix})_x=0$, for each $i<j$, so the expression in (\ref{KdV-f2-B1-f2}) really is a constant. Therefore, (\ref{KdV-Lf-kf/f4}) is a solution to our problem.
We then have
\begin{gather}\label{KdV-deltaLn-phii-2}
{\mathcal L}_n = H_n + \frac{1}{2} \sum_i\varphi_{ix}^2 -\frac{1}{2} u_0 \varphi^2-\frac{\beta}{\varphi^2}.
\end{gather}
It can be seen that this Lagrangian is rotationally invariant in the $\varphi$ space and that the expressions $(\varphi_i \varphi_{jx}-\varphi_j \varphi_{ix})$ are just the {\em angular momenta}, which are {\em constants of the motion}.
\end{subequations}

\subsection[The t\_1 flow]{The $\boldsymbol{t_1}$ flow}\label{sec:KdV-stat-t1}

We build 3 different Hamiltonians.

{\it The Lagrangian \eqref{KdV-Lagn+1},} using $8 H_2$ and labelling $-{\mathcal L}_2$ as $\mathcal L$, gives
\begin{gather}\label{L2-t1}
{\mathcal L} = \frac{1}{2} u_{0x}^2-u_0^3+\alpha u_0 \qquad\Rightarrow\quad h^{(q)} = \frac{1}{2} p_1^2+q_1^3 -\alpha q_1.
\end{gather}

{\it The Lagrangians \eqref{KdV-deltaLn-phii-1} and \eqref{KdV-deltaLn-phii-2}} are degenerate, since $H_1$ is independent of $u_{0x}$. Choosing $-H_1$ in (\ref{KdV-deltaLn-phii-1}) gives
\begin{subequations}
\begin{gather}\label{L1-t1}
\tilde {\mathcal L} = \frac{1}{2} \sum_{i=1}^N \left(\varphi_{ix}^2-u_0 \varphi_i^2 -\frac{2\beta_i}{\varphi_i^2}\right)-\frac{1}{4} u_0^2.
\end{gather}
We find $\delta_{u_0} \tilde {\mathcal L} = -\frac{1}{2} \big(\sum_{i=1}^N\varphi_i^2+u_0\big)=0$, so substitute $u_0=-\sum_{i=1}^N\varphi_i^2$, to obtain the Garnier system (see~\cite{85-8})
\begin{gather}
\tilde {\mathcal L} = \frac{1}{2} \sum_{i=1}^N \varphi_{x}^2+\frac{1}{4} \left(\sum_{i=1}^N\varphi_i^2\right)^2-\sum_{i=1}^N \frac{\beta_i}{\varphi_i^2} \nonumber\\
\qquad{} \Rightarrow\quad
 h^{(Q)} = \frac{1}{2} \sum_{i=1}^N \left(P_i^2+\frac{2\beta_i}{Q_i^2}\right)-\frac{1}{4} \left(\sum_{i=1}^N Q_i^2\right)^2.\label{L1-t1-phi}
\end{gather}

A similar calculation for the case (\ref{KdV-deltaLn-phii-2}) leads to
\begin{gather}\label{L1-t1-phi-2}
 h^{(Q)} = \frac{1}{2} \sum_{i=1}^N P_i^2 -\frac{1}{4} \left(\sum_{i=1}^N Q_i^2\right)^2 + \frac{\beta}{{\bf Q}^2},
\end{gather}
where ${\bf Q}^2 = \sum_{i=1}^N Q_i^2$.
\end{subequations}

{\bf First integrals for the Hamiltonian $\boldsymbol{h^{(Q)}}$ of (\ref{L1-t1-phi}).}
The fluxes (\ref{F12m}) give us two integrals for this Hamiltonian, one of which is $h^{(Q)}$ itself. The other is a deformation of the rotational Casimir in $N$ dimensions. To describe this, we define
\begin{subequations}
\begin{gather}\label{KdV-stat-t1-hij}
h_{ij} = (Q_iP_j-Q_jP_i)^2 +2 \left(\frac{\beta_i Q_j^2}{Q_i^2}+\frac{\beta_j Q_i^2}{Q_j^2}\right) \qquad\text{for}\quad 1\leq i <j \leq N.
\end{gather}
We then have
\begin{gather}\label{KdV-stat-t1-Fij}
F_{10} = -2 h^{(Q)}, \qquad F_{11} = \Phi^{(Q)} = \frac{1}{2} \sum_{i<j} h_{ij} + \sum_{i=1}^N \beta_i.
\end{gather}
We see from (\ref{KdV-stat-t1-Fij}) that $h_{12}$ is an integral when $N=2$. In fact, for arbitrary $N$, {\em each} $h_{ij}$ is a~first integral, for all $i<j$, so for $N\geq 3$, this system is superintegrable, as is known \cite{09-10,85-8}.
\end{subequations}

The Poisson relations for $h_{ij}$ are particularly simple when $N=3$. We first define a cubic element $h_{123}$:
 \begin{gather*}
h_{123} =(Q_1P_2-Q_2P_1)(Q_2P_3-Q_3P_2)(Q_3P_1-Q_1P_3)\\
 \hphantom{h_{123} =}{}-2\left( \!\frac{\beta_1 Q_2 Q_3 (Q_2 P_3\!-Q_3 P_2)}{Q_1^2}+ \frac{\beta_2 Q_1 Q_3 (Q_3 P_1\!-Q_1 P_3)}{Q_2^2} + \frac{\beta_3 Q_1 Q_2 (Q_1 P_2\!-Q_2 P_1)}{Q_3^2}\!\right)\!.
\end{gather*}
We then have
\begin{gather*}
\{h_{12},h_{23}\}=-\{h_{12},h_{13}\}= -\{h_{13},h_{23}\}= 4 h_{123},
\end{gather*}
and
\begin{gather*}
\{h_{12},h_{123}\} = 2\left(h_{12}(h_{13}-h_{23})-4 \beta_1 h_{23}+4 \beta_2 h_{13}\right), \\
\{h_{23},h_{123}\} = 2\left(h_{23}(h_{12}-h_{13})-4 \beta_2 h_{13}+4 \beta_3 h_{12}\right), \\
\{h_{13},h_{123}\} = 2\left(h_{13}(h_{23}-h_{12})-4 \beta_3 h_{12}+4 \beta_1 h_{23}\right). 
\end{gather*}

\begin{Remark}[rotationally symmetric case]
The Hamiltonian $h^{(Q)}$ of (\ref{L1-t1-phi-2}) is clearly {\em rotationally invariant}, so the integrals (\ref{KdV-stat-t1-hij}) are replaced by the angular momenta and $\Phi^{(Q)}$ just gives the rotational Casimir in $N$ dimensions.
\end{Remark}

\subsubsection[Bi-Hamiltonian formulation when N=1]{Bi-Hamiltonian formulation when $\boldsymbol{N=1}$}\label{sec:biHam-KdV-t1}

Here we have
\begin{gather}\label{KdV-f2-N=1}
 h^{(Q)} = \frac{1}{2} P_1^2-\frac{1}{4} Q_1^4+\frac{\beta}{Q_1^2}.
\end{gather}
The Hamiltonian formulations (\ref{L2-t1}) and (\ref{KdV-f2-N=1}) give coordinates ${\bf q}\!=\!(q_1,p_1,\alpha)$ and ${\bf Q}\!=\!(Q_1,P_1,\beta)$, respectively. From the definitions of the canonical coordinates, we have
\begin{gather}\label{PBmap-KdVt1}
q_1=u_0 =-Q_1^2 ,\qquad p_1=u_{0x} = -2 Q_1 P_1,\qquad \alpha = u_{0xx}+3 u_0^2 = -4 h^{(Q)}.
\end{gather}
In the 3D space with coordinates $\bf q$, we introduce the degenerate extension of the canonical Poisson tensor:
\begin{gather*}
{\mathcal P}_0^{(q)} = \left(
 \begin{matrix}
\phantom{-} 0 & 1 & 0 \\
 -1 & 0 & 0 \\
 \phantom{-} 0 & 0 & 0
 \end{matrix}
 \right), \qquad\text{with} \quad \frac{{\rm d}{\bf q}}{{\rm d}t_f} = {\mathcal P}_0^{(q)} \nabla_q f,
\end{gather*}
for any function $f({\bf q})$. The formulae (\ref{PBmap-KdVt1}) represent a mapping from the $\bf Q$ space to the~$\bf q$~space, with Jacobian $\frac{\pa {\bf q}}{\pa {\bf Q}}$. We use the inverse map to construct a Poisson tensor ${\mathcal P}_0^{(Q)}$ in the~$\bf Q$~space. The Jacobian of this inverse is just the inverse matrix of $\frac{\pa {\bf q}}{\pa {\bf Q}}$, which is conveniently written in terms of the $\bf Q$ coordinates:
\begin{gather*}
{\mathcal P}_0^{(Q)} = \left(\frac{\pa {\bf q}}{\pa {\bf Q}}\right)^{-1} {\mathcal P}_0^{(q)} \left(\left(\frac{\pa {\bf q}}{\pa {\bf Q}}\right)^{-1}\right)^{\rm T} ,
 \qquad\text{with}\quad \frac{{\rm d}{\bf Q}}{{\rm d}t_f} = {\mathcal P}_0^{(Q)} \nabla_Q \tilde f,
\end{gather*}
where $\tilde f ({\bf Q} ) = f ({\bf q}({\bf Q}) )$. From~(\ref{PBmap-KdVt1}), we find (up to numerical factor)
\begin{gather*}
{\mathcal P}_0^{(Q)} = \left(
 \begin{matrix}
 \phantom{-}0 & \frac{1}{Q_1^2} & -P_1 \\
 -\frac{1}{Q_1^2} & 0 & - \frac{2 \beta+Q_1^6}{Q_1^3} \\
 \phantom{-}P_1 & \frac{2 \beta+Q_1^6}{Q_1^3} & \phantom{-}0
 \end{matrix}
 \right),\qquad\text{with}\quad \alpha=-4 h^{(Q)},\quad h^{(q)} = -4 \beta.
\end{gather*}
This is compatible with the canonical bracket on the ${\bf Q}$ space:
\begin{gather*}
{\mathcal P}_1^{(Q)} = \left(
 \begin{matrix}
 \phantom{-}0 & 1 & 0 \\
 -1 & 0 & 0 \\
 \phantom{-}0 & 0 & 0
 \end{matrix}
 \right),
\end{gather*}
which in turn allows us to construct a Poisson tensor ${\mathcal P}_1^{(q)}$, using \smash{$\big(\frac{\pa {\bf q}}{\pa {\bf Q}}\big) {\mathcal P}_1^{(Q)} \big(\frac{\pa {\bf q}}{\pa {\bf Q}}\big)^{\rm T}=-4 {\mathcal P}_1^{(q)}$},
which should be written in terms of the $\bf q$ variables:
\begin{gather*}
{\mathcal P}_1^{(q)} = \left(
 \begin{matrix}
 0 & q_1 & p_1 \\
 -q_1 & 0 & \alpha-3 q_1^2 \\
 -p_1 & 3 q_1^2 - \alpha & 0
 \end{matrix}
 \right) .
\end{gather*}

We use the two independent fluxes from (\ref{F12m}):
\begin{subequations}
\begin{gather*}
F_{10} = \frac{1}{2}\alpha = -2 h^{(Q)},\qquad F_{11} = -\frac{1}{4} h^{(q)} = \beta,
\end{gather*}
to build the {\em bi-Hamiltonian ladders}, which satisfy
\begin{gather*}
 {\mathcal P}_0^{(q)}\nabla_q \alpha= 0, \qquad {\bf q}_{t_h} = {\mathcal P}_1^{(q)}\nabla_q \alpha={\mathcal P}_0^{(q)}\nabla_q h^{(q)}, \qquad {\mathcal P}_1^{(q)}\nabla_q h^{(q)} = 0, \\
 {\mathcal P}_0^{(Q)}\nabla_Q h^{(Q)}= 0, \qquad {\bf Q}_{t_h} = {\mathcal P}_1^{(Q)}\nabla_Q h^{(Q)}={\mathcal P}_0^{(Q)}\nabla_Q (-\beta), \qquad {\mathcal P}_1^{(Q)}\nabla_Q \beta = 0. 
\end{gather*}
Whereas $\alpha$ and $\beta$ were chosen to be the Casimirs of the canonical brackets, we see that $h^{(q)}$ and~$h^{(Q)}$ are the Casimirs of the ``second'' brackets.
\end{subequations}

\subsection[The t\_2 flow]{The $\boldsymbol{t_2}$ flow}\label{sec:KdV-stat-t2}

We build 3 different Hamiltonians.

{\it The Lagrangian \eqref{KdV-Lagn+1}}, with $32 H_3$ and labelling ${\mathcal L}_3$ as $\mathcal L$, gives
\begin{gather}
{\mathcal L} = \frac{1}{2} \left(5 u_0^4-10 u_0 u_{0x}^2+u_{0xx}^2\right)-\alpha u_0 \nonumber\\ \qquad\Rightarrow\quad h^{(q)} = \frac{1}{2} p_2^2+q_2 p_1 +5 q_1 q_2^2 -\frac{5}{2} q_1^4+\alpha q_1,\label{L3-t2}
\end{gather}
where we have used the generalised Legendre transformation
\begin{gather}
q_1=u_0,\qquad q_2 =u_{0x}, \qquad p_1 =\frac{\pa {\mathcal L}}{\pa u_{0x}} -\frac{{\rm d}}{{\rm d}x}\left(\frac{\pa {\mathcal L}}{\pa u_{0xx}}\right)=-10 u_0 u_{0x}-u_{0xxx}, \nonumber\\ p_2 = \frac{\pa {\mathcal L}}{\pa u_{0xx}}=u_{0xx}.\label{q1=u0}
\end{gather}

{\it The Lagrangian \eqref{KdV-deltaLn-phii-1}}, with $-8 H_2$, gives
\begin{subequations}
\begin{gather}\label{L2-f}
\tilde {\mathcal L} = \frac{1}{2} \left(u_{0x}^2+\sum_{i=1}^{N-1}\varphi_{ix}^2\right) -u_0^3-\frac{1}{2} u_0 \sum_{i=1}^{N-1}\varphi_i^2-\sum_{i=1}^{N-1} \frac{\beta_{i+1}}{\varphi_i^2}.
\end{gather}
Defining $Q_1=u_0$, $P_1=u_{0x}$, $Q_{i+1}=\varphi_i$, $P_{i+1} = \varphi_{ix}$, $i=1,\dots , N-1$, we obtain a generalised {\em H\'enon--Heiles system}
\begin{gather}\label{L2-ht}
h^{(Q)} = \frac{1}{2} \sum_{i=1}^N P_i^2 + \frac{1}{2}Q_1 \left(2 Q_1^2+ \sum_{i=2}^N Q_i^2\right)+\sum_{i=2}^N \frac{\beta_i}{Q_i^2}.
\end{gather}
The fluxes (\ref{F12m}) then give three integrals
\begin{gather}
 4 F_{20} = - h^{(Q)},\qquad 16 F_{21} = f^{(Q)} = P_1 \sum_{i=1}^N Q_i P_i -2 Q_1 h^{(Q)} + \frac{1}{8} \big(16 Q_1^4+12 Q_1^2 {\bf Q}^2+{\bf Q}^4\big),\nonumber\\
 64 F_{22} = \Phi^{(Q)} = \frac{1}{2} \sum_{2\leq i<j\leq N} h_{ij} + \sum_{i=2}^N \beta_i,\label{KdV-stat-t2-Fij}
\end{gather}
where $h_{ij}$ are given by (\ref{KdV-stat-t1-hij}) and ${\bf Q}^2 = \sum_{i=2}^N Q_i^2$. Again, {\em each} $h_{ij}$ is a first integral.
\end{subequations}

{\it For the case \eqref{KdV-deltaLn-phii-2}}, an identical calculation leads to
\begin{gather*}
 h^{(Q)} = \frac{1}{2} \sum_{i=1}^N P_i^2 + \frac{1}{2}Q_1 \left(2 Q_1^2+ \sum_{i=2}^N Q_i^2\right)+ \frac{\beta}{{\bf Q}^2}.
\end{gather*}
This Hamiltonian is clearly {\em rotationally invariant} (in 
$Q_2,\dots ,Q_N$ space), so the integrals~(\ref{KdV-stat-t1-hij}) are replaced by the angular momenta and $F_{22}$ just gives the rotational Casimir in $(N-1)$ dimensions.

\subsubsection[Bi-Hamiltonian formulation when N=2]{Bi-Hamiltonian formulation when $\boldsymbol{N=2}$}\label{sec:biHam-KdV-t2}

Here we have the usual 2 degrees of freedom H\'enon--Heiles Hamiltonian
\begin{gather}\label{L2-ht-N=2}
h^{(Q)} = \frac{1}{2} \big(P_1^2+P_2^2\big) + \frac{1}{2}Q_1 \big(2 Q_1^2+ Q_2^2\big)+ \frac{\beta}{Q_2^2}.
\end{gather}
The Hamiltonian formulations (\ref{L3-t2}) and (\ref{L2-ht-N=2}) gave coordinates ${\bf q} = (q_i,p_i,\alpha)$ and ${\bf Q} = (Q_i,P_i,\beta)$, respectively. From the definitions of the canonical coordinates, we have (\ref{q1=u0}) and $\alpha = 10 u_0^3+5 u_{0x}^2+10 u_0 u_{0xx}+u_{0xxxx}$.
Since $u_0 = Q_1$, we can use $h^{(Q)}$ to calculate all $x$-derivatives: $u_0^{(m+1)} = \big\{u_0^{(m)}, h^{(Q)} \big\}$, where $u_0^{(i)}$ is the $i^{\rm th}$ derivative of $u_0$ with respect to $x$, giving
\begin{gather}
 q_1=Q_1,\qquad q_2=P_1,\qquad p_1=-4Q_1P_1+Q_2P_2,\qquad p_2=-3Q_1^2-\frac12Q_2^2, \nonumber\\ \alpha = -2 h^{(Q)}.\label{q1=Q1}
\end{gather}
The three independent fluxes from (\ref{F12m}) give
\begin{subequations}
\begin{gather}\label{KdVt2-F2k}
4 F_{20} = \frac{1}{2}\alpha = - h^{(Q)},\qquad 16 F_{21} = h^{(q)} = f^{(Q)},\qquad 64 F_{22} = -\frac{1}{2} f^{(q)} = \beta,
\end{gather}
where
\begin{gather}
f^{(q)} = p_1^2+4 q_1 p_2^2+8 q_1q_2p_1 + 2 \big(10 q_1^3-q_2^2\big) p_2 +2 q_1^2 \big(12 q_1^3+5 q_2^2\big) - 2 \alpha \big(p_2+3 q_1^2\big), \label{KdVt2-f} \\
f^{(Q)} = P_2 (Q_2 P_1-Q_1 P_2) + \frac{1}{8} Q_2^2 \big(4 Q_1^2+Q_2^2\big)-\frac{2\beta Q_1}{Q_2^2}. \label{KdVt2-ft}
\end{gather}
\end{subequations}

In the 5D space with coordinates $\bf q$, we introduce a degenerate extension of the canonical Poisson tensor and use (\ref{q1=Q1}) as a Poisson map to obtain ${\mathcal{P}}_0^{(Q)}$:
 \begin{gather*}
 {\mathcal{P}}_0^{(q)}= \left( \begin{matrix}
 \phantom{-}0& \phantom{-}0&1&0&0 \\
 \phantom{-}0& \phantom{-}0&0&1&0 \\
 -1& \phantom{-}0&0&0&0 \\
 \phantom{-} 0&-1&0&0&0 \\
 \phantom{-} 0& \phantom{-}0&0&0&0
 \end{matrix}\right)
 \quad \Rightarrow\quad
{\mathcal{P}}_0^{(Q)} = \left( \begin{matrix}
 \phantom{-}0& \phantom{-}0& \phantom{-}0& \phantom{-}\frac1{Q_2}& a_{15} \\[2mm]
 \phantom{-} 0& \phantom{-}0& \phantom{-}\frac1{Q_2}&-\frac{2Q_1}{Q_2^2}& a_{25} \\[2mm]
 \phantom{-} 0&-\frac1{Q_2}& \phantom{-}0& \phantom{-}\frac{P_2}{Q_2^2}& a_{35} \\[2mm]
 -\frac1{Q_2}&\frac{2Q_1}{Q_2^2}&-\frac{P_2}{Q_2^2}& \phantom{-}0& a_{45} \\[2mm]
 -a_{15} & -a_{25} &-a_{35} & -a_{45} &0
 \end{matrix}\right),
\end{gather*}
where the column $(a_{15},a_{25},a_{35},a_{45},0)^{\rm T}=-P_1^{(Q)}\nabla_Q f^{(Q)}$, with ${\mathcal{P}}_1^{(Q)}$ being the compatible canonical bracket on the $\bf Q$ space, which is similarly transformed to the $\bf q$ space:
 \begin{gather*}
 {\mathcal{P}}_1^{(Q)}= \begin{pmatrix}
 \phantom{-}0& \phantom{-}0&1&0&0 \\
 \phantom{-}0& \phantom{-}0&0&1&0 \\
 -1& \phantom{-}0&0&0&0 \\
 \phantom{-}0&-1&0&0&0 \\
 \phantom{-} 0& \phantom{-}0&0&0&0
 \end{pmatrix} \\
 \qquad\Rightarrow\quad
{\mathcal{P}}_1^{(q)} =\begin{pmatrix}
 \phantom{-}0 & 1 & -4q_1 & \phantom{-}0 & b_{15} \\
 -1&0& \phantom{-}4q_2& \phantom{-}6q_1& b_{25} \\
 \phantom{-}4q_1&-4q_2& \phantom{-}0&-30q_1^2-2p_2& b_{35} \\
 \phantom{-}0&-6q_1&30q_1^2+2p_2& \phantom{-}0& b_{45} \\
 -b_{15} & -b_{25} & -b_{35} &-b_{45} &0
 \end{pmatrix},
\end{gather*}
where the column $(b_{15},b_{25},b_{35},b_{45},0)^{\rm T}=-2P_0^{(q)}\nabla_q h^{(q)}$.

The Hamiltonian ladders are
\begin{gather*}
 {\mathcal P}_0^{(q)}\nabla_q \alpha=0, \qquad {\mathcal P}_1^{(q)}\nabla_q\bigl(-\tfrac{\alpha}{2}\bigr)={\mathcal P}_0^{(q)}\nabla_q h^{(q)},
\nonumber\\ {\mathcal P}_1^{(q)}\nabla_q h^{(q)} = {\mathcal P}_0^{(q)}\nabla_q \big(\tfrac{1}{2} f^{(q)}\big), \qquad {\mathcal P}_1^{(q)}\nabla_q f^{(q)} = 0, \\ 
{\mathcal{P}}_0^{(Q)}\nabla_{Q} h^{(Q)} =0,\qquad
 {\mathcal{P}}_1^{(Q)}\nabla_{Q} h^{(Q)} = {\mathcal{P}}_0^{(Q)}\nabla_{Q} f^{(Q)},\nonumber\\
 {\mathcal{P}}_1^{(Q)}\nabla_{Q} f^{(Q)} ={\mathcal{P}}_0^{(Q)}\nabla_{Q}(-\beta),\qquad
 {\mathcal{P}}_1^{(Q)}\nabla_{Q}\beta=0. 
\end{gather*}

The two nontrivial flows correspond to the two lower commuting flows of the KdV hierarchy. We see this by looking at the $u_0$ ($=q_1=Q_1$) component of the flows.

In the $\bf q$ space, the flows ${\mathcal P}_0^{(q)}\nabla_q h^{(q)}$ and ${\mathcal P}_0^{(q)}\nabla_q \bigl(-\frac{1}{2} f^{(q)}\bigr)$ give
\begin{subequations}
\begin{gather}\label{KdVt2-qthtf}
q_{1t_h} = \frac{\pa h^{(q)}}{\pa p_1} = q_2 = q_{1x},\qquad q_{1t_f} = \frac{\pa}{\pa p_1} \bigl(-\tfrac{1}{2} f^{(q)}\bigr) = -p_1-4 q_1q_2 = q_{1xxx}+6q_1 q_{1x}.\!\!\!
\end{gather}
In the $\bf Q$ space, the flows ${\mathcal P}_1^{(Q)}\nabla_Q h^{(Q)}$ and ${\mathcal P}_1^{(Q)}\nabla_Q \left(-f^{(Q)}\right)$ give
\begin{gather}\label{KdVt2-Qthtf}
Q_{1t_{h}} = \frac{\pa h^{(Q)}}{\pa P_1} = P_1 = Q_{1x},\qquad Q_{1t_{f}} = -\frac{\pa f^{(Q)}}{\pa P_1} = -Q_2 P_2 = \big(Q_{1xx}+3 Q_1^2\big)_x.
\end{gather}
\end{subequations}

\subsection{The Lax representation}\label{sec:sKdV-Lax}

As we previously said, to obtain a Lax representation for stationary flows, we use the {\em zero curvature} representation, given in (\ref{zero-c}) and (\ref{Utn}), with $U_{t_n}=0$. When written in terms of the coordinates on the stationary manifold, the characteristic equation, $\det\big(z I - L^{(n)}\big)=0$, is $\lambda$-dependent, giving the same integrals as derived from the fluxes~$F_{ij}$.

\subsubsection[The t\_1 flow]{The $\boldsymbol{t_1}$ flow}

In this case our matrices are
\begin{gather}\label{UV1}
U = \left(
 \begin{matrix}
 0 & 1 \\
 \lambda-u_0 & 0
 \end{matrix}
 \right),\qquad V^{(1)} = \frac{1}{2}\left(
 \begin{matrix}
 -u_{0x} & 4\lambda+2 u_0 \\
 4 \lambda^2-2\lambda u_0-2u_0^2-u_{0xx} & u_{0x}
 \end{matrix}
 \right).
\end{gather}

{\it Using the coordinates of \eqref{L2-t1}}, we define
\begin{gather*}
L^{(1)} = -2 V^{(1)} =\begin{pmatrix}
 p_1 & -2 (2\lambda + q_1) \\
 \alpha-q_1^2+2 \lambda q_1-4 \lambda^2 & -p_1 \\
 \end{pmatrix} \nonumber\\
 \qquad\Rightarrow\quad z^2-2 h^{(q)}+4 \alpha \lambda -16 \lambda^3=0,
\end{gather*}
where $\alpha$ and $h^{(q)}$ are constants of the motion. The Lax representation of the equation generated by $h^{(q)}$ is
\begin{gather*}
L^{(1)}_x = \big\{L^{(1)},h^{(q)}\big\} = \big[U,L^{(1)}\big] ,
\end{gather*}
using the canonical Poisson bracket, acting componentwise on the matrix.

{\it Using the coordinates of \eqref{L1-t1-phi}}, we have
\begin{subequations}
\begin{gather}
L^{(1)} = V^{(1)} =\begin{pmatrix}
 \displaystyle\sum_{1}^N Q_iP_i & 2\lambda -{\bf Q}^2 \\
 \displaystyle\sum_{1}^N\!\left(P_i^2+\frac{2\beta_i}{Q_i^2}\right)\!+\lambda \big(2 \lambda+{\bf Q}^2\big) & - \displaystyle\sum_{1}^N Q_iP_i \\
 \end{pmatrix} \nonumber\\ \qquad\Rightarrow\quad z^2-4 \lambda h^{(Q)}\!+ 2 \Phi^{(Q)}\!-4 \lambda^3=0,\label{L1-phi}
\end{gather}
where ${\bf Q}^2=\sum_{i=1}^N Q_i^2$, with $h^{(Q)}$ and $\Phi^{(Q)}$ as given in (\ref{L1-t1-phi}) and (\ref{KdV-stat-t1-Fij}). The Lax equations generated by $h^{(Q)}$ and $\Phi^{(Q)}$ are
\begin{gather}\label{L1x-phi}
L^{(1)}_x = \big\{L^{(1)},h^{(Q)}\big\} = \big[U,L^{(1)}\big] , \qquad \big\{L^{(1)}, \Phi^{(Q)}\big\} = 0.
\end{gather}
\end{subequations}
In fact, $\big\{L^{(1)}, h_{ij}\big\} = 0$, for each of the functions $h_{ij}$ of~(\ref{KdV-stat-t1-hij}).

When $N=1$, the function $\Phi^{(Q)}$ is replaced by $\beta$.

\begin{Remark}[rotationally symmetric case]
Similar equations hold for the case (\ref{L1-t1-phi-2}), with $\sum_{1}^N \frac{\beta_i}{Q_i^2}$ replaced by $\frac{\beta}{{\bf Q}^2}$ and the integral $\Phi^{(Q)}$ by the rotational Casimir.
\end{Remark}

\subsubsection[The t\_2 flow]{The $\boldsymbol{t_2}$ flow}

In this case our principal matrix is proportional to $V^{(2)}$, which is to be written in terms of appropriate coordinates.

{\it Using the coordinates from \eqref{L3-t2}}, we define
\begin{gather*}
L^{(2)} = -8 V^{(2)} = \left(
 \begin{matrix}
 4 \lambda q_2-p_1-4 q_1 q_2 & -2 \big(8\lambda^2 +4 \lambda q_1+p_2+3 q_1^2\big) \\
 a_{21} & p_1+4 q_1 q_2 - 4 \lambda q_2
 \end{matrix}
 \right) ,
\end{gather*}
where $a_{21} = \alpha-2 q_1 p_2-4 q_1^3+q_2^2+2 \lambda \big(p_2+q_1^2\big)+8 q_1 \lambda^2-16 \lambda^3$. The characteristic equation is
\begin{gather*}
z^2 -256 \lambda^5+16 \alpha \lambda^2 +8\lambda h^{(q)}-f^{(q)} =0,
\end{gather*}
with $h^{(q)}$ and $f^{(q)}$ given by (\ref{L3-t2}) and (\ref{KdVt2-f}). We have
\begin{gather*}
L^{(2)}_x = \big\{L^{(2)},h^{(q)}\big\} = \big[U,L^{(2)}\big],\qquad L^{(2)}_{t_1} = \big\{L^{(2)},f^{(q)}\big\} = \big[L^{(1)},L^{(2)}\big],
\end{gather*}
with
\begin{gather*}
U = \left(
 \begin{matrix}
 0 & 1 \\
 \lambda-q_1 & 0
 \end{matrix}
 \right),\qquad L^{(1)} = \left(
 \begin{matrix}
 2 q_2 & -4 (2 \lambda+q_1) \\
 2\big(p_2+2 q_1^2+2\lambda q_1-4\lambda^2\big) & -2 q_2
 \end{matrix}
 \right),
\end{gather*}
where $L^{(1)}$ is derived from $V^{(1)}$ of (\ref{UV1}).

{\it Using the coordinates of \eqref{L2-ht}}, we define
\begin{subequations}
\begin{gather}\label{L2-Lax-ht}
L^{(2)} = 8 V^{(2)} = \left(
 \begin{matrix}
 \displaystyle\sum_2^N Q_i P_i - 4 \lambda P_1 & 16\lambda^2 + 8 \lambda Q_1-{\bf Q}^2 \\
 a_{21} & \displaystyle4 \lambda P_1 - \sum_2^N Q_i P_i
 \end{matrix}
 \right) ,
\end{gather}
where
\begin{gather*} {\bf Q}^2=\sum_2^N Q_i^2, \qquad a_{21}=16 \lambda^3-8 Q_1 \lambda^2+\lambda \left(4 Q_1^2+\sum_2^N Q_i^2\right)+\sum_2^N \left(P_i^2+\frac{2 \beta_i}{Q_i^2}\right).
\end{gather*}
 The characteristic equation is
\begin{gather}\label{L2-Lax-ht-char}
z^2=256 \lambda^5+32 h^{(Q)} \lambda^2-8 f^{(Q)} \lambda -2 \Phi^{(Q)},
\end{gather}
where $f^{(Q)}$ and $\Phi^{(Q)}$ are given in (\ref{KdV-stat-t2-Fij}). We have
\begin{gather}
L^{(2)}_x = \big\{L^{(2)},h^{(Q)} \big\} = \big[U,L^{(2)} \big],\qquad L^{(2)}_{t_1} = \big\{L^{(2)},f^{(Q)} \big\} = \big[L^{(1)},L^{(2)} \big], \nonumber\\ \big\{L^{(2)},h_{ij} \big\}=0,\label{L2-laxeq-Q}
\end{gather}
for $2\leq i<j\leq N$, and where
\begin{gather}\label{L2-laxmats-Q}
U = \left(
 \begin{matrix}
 0 & 1 \\
 \lambda-Q_1 & 0
 \end{matrix}
 \right),\qquad L^{(1)} = \left(
 \begin{matrix}
 P_1 & -2 (2 \lambda+Q_1) \\
 -4\lambda^2+2 \lambda Q_1-\frac{1}{2}\big(2 Q_1^2+{\bf Q}^2\big) & -P_1
 \end{matrix}
 \right).
\end{gather}
\end{subequations}

\section{DWW stationary flows}\label{sec:DWW-stat}

We can follow the approach of Section \ref{sec:KdV-stat} in the context of the DWW hierarchy. We have three representations of ${\bf u}_{t_n}$, given by (\ref{DWW-triHam}), with a {\em stationary} flow being the reduction to a {\em finite-dimensional} space with ${\bf u}_{t_n}=0$. The ``time'' variable for this flow is $x$.
In the KdV case the Casimir of $B_0$ was just $H_0$, which gave us the $\alpha$ term in (\ref{KdV-Lagn+1}), whilst, for the DWW operator $B_0$, we have {\em two} Casimirs, $H_0$, $H_1$. The DWW operator $B_1$ has {\em one} Casimir $H_0$, corresponding to the first order part, but we also have the squared eigenfunctions for $J_0$. We leave the operator~$B_2$ to a later discussion of the Miura maps in Section \ref{sec:DWW-triH}, so now consider the two operators $B_0$ and~$B_1$:%

1.~Using $B_0$, we have
 \begin{subequations}
 \begin{gather}\label{DWW-B0dH=0}
B_0 \delta_{u} H_{n+2}=0 \quad\Rightarrow\quad \delta_{u} (H_{n+2}-\alpha_0 H_0-4\alpha_1 H_1)=0 ,
\end{gather}
giving the Lagrangian
\begin{gather}\label{DWW-Lagn+2}
{\mathcal L}_{n+2} = H_{n+2} - \alpha_0 u_1-\alpha_1 \big(4 u_0+u_1^2\big).
\end{gather}
\end{subequations}
Again, we use the (generalised) Legendre transformation to find canonical coordinates $(q_i,p_i)$ and the Hamiltonian function.

2.~Using $B_1$, we have
 \begin{gather}
\delta_{u_0} H_{n+1} = \frac{1}{2} \varphi^2,\qquad \delta_{u_1} H_{n+1} = -\alpha \nonumber\\
\qquad{} \Rightarrow\quad
 {\mathcal L}_{n+1} = H_{n+1}+\alpha u_1 + \frac{1}{2} \varphi_x^2-\frac{1}{2} u_0 \varphi^2-\frac{\beta}{\varphi^2}.\label{DWW-dH=phi2}
\end{gather}
Again, the extensions (\ref{KdV-deltaLn-phii-1}) and (\ref{KdV-deltaLn-phii-2}) can be introduced here.

\subsection[The t\_1 flow]{The $\boldsymbol{t_1}$ flow}

We build 3 different Hamiltonians.

{\it The Lagrangian \eqref{DWW-Lagn+2}}, using $8 H_3$ and labelling $-{\mathcal L}_3$ as $\mathcal L$, gives
\begin{subequations}
\begin{gather}\label{DWW-B0-t1}
{\mathcal L} = \frac{1}{2} u_{1x}^2-\frac{5}{8} u_1^4+\alpha_0 u_1+\alpha_1 u_1^2+u_0 \big(4 \alpha_1-3 u_1^2\big)-2 u_0^2 .
\end{gather}
This Lagrangian is degenerate, but $\delta_{u_0} {\mathcal L}=0$ gives $u_0=\alpha_1-\frac{3}{4} u_1^2$, after which
\begin{gather}
{\mathcal L} = \frac{1}{2} u_{1x}^2+\frac{1}{2} u_1^4-2\alpha_1 u_1^2+\alpha_0 u_1+2 \alpha_1^2 \nonumber\\
\qquad{} \Rightarrow\quad h^{(q)} = \frac{1}{2} p_1^2-\frac{1}{2} q_1^4+2\alpha_1 q_1^2-\alpha_0 q_1.\label{DWW-B0-t1-h}
\end{gather}
The fluxes (\ref{DWW-F12m}) give
\begin{gather}\label{DWW-B0-t1-F1j}
F_{10} = 2 \alpha_1,\qquad F_{11} = \frac{1}{2} \alpha_0, \qquad F_{12} = -\frac{1}{4} h^{(q)} +\frac{1}{2} \alpha_1^2.
\end{gather}
\end{subequations}

{\it The Lagrangian \eqref{DWW-dH=phi2}}, using $-H_2$ in the first multicomponent version of (\ref{DWW-dH=phi2}) (see (\ref{KdV-deltaLn-phii-1})), we find
\begin{subequations}
\begin{gather}\label{DWW-B1-t1}
\tilde {\mathcal L} = \sum_{i=1}^N \left(\frac{1}{2} \varphi_{ix}^2 -\frac{1}{2} u_0 \varphi_i^2-\frac{\beta_i}{\varphi_i^2}\right)+\alpha u_1-\frac{1}{8} u_1 \left(4 u_0+u_1^2\right).
\end{gather}
This is also degenerate, but $\delta_{u_0}\tilde {\mathcal L}=0$ and $\delta_{u_1}\tilde {\mathcal L}=0$ give $u_1=-\sum \varphi_i^2$ and $u_0=2 \alpha-\frac{3}{4} \big(\sum \varphi_i^2\big)^2$, after which
\begin{gather}\label{DWW-B1-t1-1}
\tilde {\mathcal L} = \sum_{i=1}^N \left(\frac{1}{2} \varphi_{ix}^2 -\alpha \varphi_i^2-\frac{\beta_i}{\varphi_i^2}\right)+\frac{1}{8} \left(\sum_{i=1}^N \varphi_i^2\right)^3.
\end{gather}
Defining $Q_i=\varphi_i$, $P_i = \varphi_{ix}$, we obtain the Hamiltonian
\begin{gather}\label{DWW-B1-t1-ht}
h^{(Q)} = \frac{1}{2} \sum_{i=1}^N \left(P_i^2+2\alpha Q_i^2+\frac{2\beta_i}{Q_i^2}\right)-\frac{1}{8} \left(\sum_{i=1}^N Q_i^2\right)^3 .
\end{gather}
The fluxes (\ref{DWW-F12m}) give (compare with (\ref{KdV-stat-t1-Fij}))
\begin{gather}\label{DWW-B1-t1-F1j}
F_{10} = 4 \alpha,\qquad F_{11} = -2 h^{(Q)}, \qquad F_{12}- 2\alpha^2 = \Phi^{(Q)} = \frac{1}{2}\sum_{1\leq i<j\leq N} h_{ij}+\sum_{i=1}^N \beta_i.
\end{gather}
\end{subequations}
Since $h_{ij}$ are a deformation of the (squares of) angular momenta, it is straightforward to build an involutive set of integrals, so (\ref{DWW-B1-t1-ht}) defines a {\em superintegrable system} (for $N\geq 3$).

When $N=1$, $F_{12}=2\alpha^2+\beta$.
When $N=2$, we have 2 integrals $h^{(Q)}$, $h_{12}$.
When $N=3$, we have 4 integrals $h^{(Q)}$, $h_{ij}$, with $h^{(Q)}$, $h_{12}$ and $\Phi^{(Q)}$ in involution. This and the Hamiltonian~(\ref{L1-t1-phi}) (for $N=3$) are particular cases of the first potential in Table~II of~\cite{90-22}.

\begin{Remark}[rotationally invariant case]
The extension (\ref{KdV-deltaLn-phii-2}) gives a similar result, with
\begin{gather}\label{DWW-B1-t1-ht-rot}
h^{(Q)} = \frac{1}{2} \sum_{i=1}^N \left(P_i^2+2\alpha Q_i^2\right)-\frac{1}{8} \left(\sum_{i=1}^N Q_i^2\right)^3 +\frac{\beta}{{\bf Q}^2},
\end{gather}
where ${\bf Q}^2=\sum_{i=1}^N Q_i^2$. The integrals $h_{ij}$ are just replaced by the rotation algebra.
\end{Remark}

\subsubsection[Bi-Hamiltonian formulation when N=1]{Bi-Hamiltonian formulation when $\boldsymbol{N=1}$}\label{sec:DWW-N=1biHam}

Here we have
\begin{gather}\label{DWW-f2-N=1}
 h^{(Q)} = \frac{1}{2} P_1^2-\frac{1}{8} Q_1^6+\alpha Q_1^2+\frac{\beta}{Q_1^2}.
\end{gather}
The Hamiltonian formulations (\ref{DWW-B0-t1-h}) and (\ref{DWW-f2-N=1}) give coordinates ${\bf q}=(q_1,p_1,\alpha_0,\alpha_1)$ and ${\bf Q}=(Q_1,P_1,\beta,\alpha)$, respectively.
From the definitions of the canonical coordinates, we have
\begin{gather*}
q_1 = -Q_1^2 ,\qquad p_1 = -2 Q_1 P_1, \qquad \alpha_0 = -4 h^{(Q)}, \qquad \alpha_1=2 \alpha.
\end{gather*}
Following the same procedure as Section~\ref{sec:biHam-KdV-t1}, we introduce a degenerate extension of the canonical Poisson tensor in the 4D space with coordinates $\bf q$ and then define
\smash{$\big(\frac{\pa {\bf Q}}{\pa {\bf q}}\big) {\mathcal P}_0^{(q)} \big(\frac{\pa {\bf Q}}{\pa {\bf q}}\big)^{\rm T}= \frac{1}{4} {\mathcal P}_0^{(Q)}$}, giving
\begin{gather*}
{\mathcal P}_0^{(q)} = \left(
 \begin{matrix}
 \phantom{-}0 & 1 & 0 & 0 \\
 -1 & 0 & 0 & 0 \\
 \phantom{-}0 & 0 & 0 & 0\\
 \phantom{-}0 & 0 & 0 & 0
 \end{matrix}
 \right) \qquad\text{and}\qquad
{\mathcal P}_0^{(Q)} = \left(
 \begin{matrix}
 \phantom{-}0 & \phantom{-}\frac{1}{Q_1^2} & a_{13} & 0 \\
 -\frac{1}{Q_1^2} & \phantom{-}0 & a_{23} & 0 \\
 -a_{13} & -a_{23} & 0 & 0\\
 \phantom{-}0 & \phantom{-}0 & 0 & 0
 \end{matrix}
 \right),
\end{gather*}
where \smash{$(a_{13},a_{23},0,0)^{\rm T}=-P_1^{(Q)} \nabla_Q h^{(Q)}$} and \smash{${\mathcal P}_1^{(Q)}$} is the compatible (degenerate) canonical Poisson tensor on the $\bf Q$ space, which in turn leads to
\smash{$\big(\frac{\pa {\bf q}}{\pa {\bf Q}}\big) {\mathcal P}_1^{(Q)} \big(\frac{\pa {\bf q}}{\pa {\bf Q}}\big)^{\rm T}=-4 {\mathcal P}_1^{(q)}$}, giving
\begin{gather*}
{\mathcal P}_1^{(Q)} = \left(
 \begin{matrix}
\phantom{-}0 & 1 & 0 & 0 \\
 -1 & 0 & 0 & 0 \\
 \phantom{-}0 & 0 & 0 & 0\\
 \phantom{-}0 & 0 & 0 & 0
 \end{matrix}
 \right) \qquad\text{and}\qquad
{\mathcal P}_1^{(q)} = \left(
 \begin{matrix}
 \phantom{-}0 & \phantom{-}q_1 & b_{13} & 0 \\
 -q_1 & \phantom{-}0 & b_{23} & 0 \\
 -b_{13} & -b_{23} & 0 & 0\\
 \phantom{-}0 & \phantom{-}0 & 0 & 0
 \end{matrix}
 \right),
\end{gather*}
where \smash{$(b_{13},b_{23},0,0)^{\rm T}=P_0^{(q)} \nabla_q h^{(q)}$}.

We use the three independent fluxes from (\ref{DWW-F12m}) to build the {\em bi-Hamiltonian ladders}:
\begin{gather*}
F_{10} = 2\alpha_1 = 4 \alpha,\qquad F_{11} = \frac{1}{2} \alpha_0 = -2 h^{(Q)},\qquad F_{12} = -\frac{1}{4} h^{(q)}+\frac{1}{2} \alpha_1^2 = 2 \alpha^2+ \beta,
\end{gather*}
which satisfy ${\mathcal P}_1^{(q)}\nabla_q \alpha_1= {\mathcal P}_0^{(q)}\nabla_q \alpha_1= {\mathcal P}_1^{(Q)}\nabla_Q \alpha= {\mathcal P}_0^{(Q)}\nabla_Q \alpha=0$ and
\begin{gather*}
 {\mathcal P}_0^{(q)}\nabla_q \alpha_0= 0, \qquad {\bf q}_{t_h} = {\mathcal P}_1^{(q)}\nabla_q \alpha_0={\mathcal P}_0^{(q)}\nabla_q h^{(q)}, \qquad
 {\mathcal P}_1^{(q)}\nabla_q h^{(q)} = 0, \label{DWWt1-q-flows}\\
 {\mathcal P}_0^{(Q)}\nabla_Q h^{(Q)}= 0, \qquad
 {\bf Q}_{t_h} = {\mathcal P}_1^{(Q)}\nabla_Q h^{(Q)} ={\mathcal P}_0^{(Q)}\nabla_Q (-\beta), \qquad {\mathcal P}_1^{(Q)}\nabla_Q \beta = 0. 
\end{gather*}

\begin{Remark}[the parameters $\alpha$ and $\alpha_1$]
At this stage it looks like we should consider~$\alpha_1$ and~$\alpha$ as \emph{constants} rather than dynamical variables. However, we will see in Section~\ref{sec:DWW-triH} that they generate nontrivial flows with respect to a \emph{third} Poisson bracket.
\end{Remark}

\subsection[The t\_2 flow]{The $\boldsymbol{t_2}$ flow}

We build 3 different Hamiltonians.

{\it The Lagrangian \eqref{DWW-Lagn+2}}, using $2 H_4$ and labelling ${\mathcal L}_4$ as $\mathcal L$, gives
\begin{subequations}
\begin{gather}\label{DWW-B0-t2}
{\mathcal L} = \frac{1}{64} \big(u_1\big(4 u_0 +u_1^2\big)\big(12 u_0+7 u_1^2\big)-4 u_{1x}(4 u_{0x}+5 u_1 u_{1x})\big)-\alpha_0 u_1 - \alpha_1 \big(4 u_0+u_1^2\big).\!\!\!
\end{gather}
The standard Legendre transformation gives
\[
u_0=q_1,\qquad u_1=q_2,\qquad u_{0x} =2 (5q_2p_1-2 p_2), \qquad u_{1x}=-4 p_1,
\]
giving the Hamiltonian
\begin{gather}\label{DWW-B0-t2-h}
h^{(q)} = 5 q_2 p_1^2-4 p_1 p_2-\frac{q_2}{64} \big(4 q_1+q_2^2\big)\big(12 q_1+7 q_2^2\big)+\alpha_0 q_2 +\alpha_1\big(4 q_1+q_2^2\big) .
\end{gather}
The fluxes (\ref{DWW-F12m}) give
\begin{gather}
 F_{20} = 8 \alpha_1,\qquad F_{21} = 2 \alpha_0, \qquad F_{22} = h^{(q)},\label{DWW-B0-t2-F2j}\\
 F_{23}-8 \alpha_1^2 = f^{(q)} = \frac{1}{256} \big(4 q_1+3 q_2^2\big) \big(64(\alpha_0-2 \alpha_1 q_2)+\!64 p_1^2-16 q_1^2+5 q_2^4\big)-2 (p_2-q_2 p_1)^2.\nonumber
\end{gather}
\end{subequations}

{\it The Lagrangian \eqref{DWW-dH=phi2}}, using $-8H_3$ in the first multicomponent version of (\ref{DWW-dH=phi2}) (see (\ref{KdV-deltaLn-phii-1})), we find
\begin{subequations}
\begin{gather*}
\tilde {\mathcal L} = \sum_{i=1}^{N-1} \left(\frac{1}{2} \varphi_{ix}^2 -\frac{1}{2} u_0 \varphi_i^2-\frac{\beta_{i+1}}{\varphi_i^2}\right)+\alpha u_1+\frac{1}{2} u_{1x}^2-\frac{1}{8} \big(4 u_0+u_1^2\big)\big(4 u_0+5 u_1^2\big).
\end{gather*}
This is degenerate in $u_0$, but $\delta_{u_0}\tilde {\mathcal L}=0$ gives $u_0=-\frac{3}{4} u_1^2-\frac{1}{8} \sum_1^{N-1} \varphi_i^2$, after which
\begin{gather*}
\tilde {\mathcal L} = \frac{1}{2} \left(u_{1x}^2+\sum_{i=1}^{N-1} \left(\varphi_{ix}^2-\frac{2\beta_{i+1}}{\varphi_i^2}\right)\right) +\alpha u_1
 +\frac{1}{32} \left(16 u_1^4+12 u_1^2\sum_{i=1}^{N-1} \varphi_i^2 + \left(\sum_{i=1}^{N-1} \varphi_i^2\right)^2\right).
\end{gather*}
Defining $u_1=Q_1$, $u_{1x}=P_1$ and $\varphi_{i}=Q_{i+1}$, $\varphi_{ix}=P_{i+1}$, $i=1, \dots , N-1$, we obtain the Hamiltonian
\begin{gather}\label{DWW-B1-t2-ht}
h^{(Q)} = \frac{1}{2} \sum_{i=1}^N P_i^2-\alpha Q_1+\sum_{i=2}^N\frac{\beta_i}{Q_i^2}-\frac{1}{32} \left(16 Q_1^4+12 Q_1^2\sum_{i=2}^N Q_i^2+ \left(\sum_{i=2}^N Q_i^2\right)^2\right) .
\end{gather}
The fluxes (\ref{DWW-F12m}) give
\begin{gather}
 F_{20} = \frac{1}{2} \alpha,\qquad F_{21} = -\frac{1}{4} h^{(Q)},\qquad 64 F_{23}-2 \alpha^2 = \Phi^{(Q)} = \frac{1}{2} \sum_{2\leq i<j\leq N} h_{ij} + \sum_{i=2}^N \beta_i, \label{DWW-B1-t2-F1j}\\
 16 F_{22}= f^{(Q)} = P_1 \sum_i^N Q_i P_i-2 Q_1 h^{(Q)}-\frac{1}{2}\alpha\!\big(4 Q_1^2+{\bf Q}^2\big) -\frac{1}{16} Q_1\big(4 Q_1^2+ {\bf Q}^2\big) \big(4 Q_1^2+ 3{\bf Q}^2\big),\nonumber
\end{gather}
where ${\bf Q}^2=\sum_2^N Q_i^2$ and $h_{ij}$ are defined by (\ref{KdV-stat-t1-hij}), with $2\leq i<j\leq N$.
\end{subequations}
In fact, $h_{ij}$ are themselves first integrals.
Since $h_{ij}$ are a deformation of the (squares of) angular momenta, it is straightforward to build an involutive set of integrals, so (\ref{DWW-B1-t2-ht}) defines a {\em superintegrable system} (for $N\geq 4$).

When $N=2$, $F_{23}= \frac{1}{64}\big(2\alpha^2+\beta\big)$, so there are just the two integrals, $h^{(Q)}$ and $f^{(Q)}$. This is the case which was found in the classifications of \cite{83-12,82-8}. We show that this particular case is {\em tri-Hamiltonian} in Section \ref{sec:DWW-triH} (see Remark~\ref{Rem:3HamQuartic}). In fact, this Hamiltonian and the H\'enon--Heiles one~(\ref{L2-ht-N=2}) belong to an infinite family, which are separable in parabolic coordinates (see, for example, \cite[equation~(2.2.41)]{90-16}). Other members of the classifications of~\cite{83-12,82-8} can similarly be derived as stationary flows associated with $4^{\rm th}$ order scalar Lax operators~\cite{f95-3}.

When $N=3$, we have $h^{(Q)}$, $f^{(Q)}$ and $\Phi^{(Q)}$ in involution. It appears in the classification of~\cite[case~4 of Table~1]{86-8} and is further generalised in~\cite{f22-1} and in Section~\ref{sec:generalisations} of this paper.

When $N=4$, we have $h^{(Q)}$, $f^{(Q)}$, $h_{23}$ and $\Phi^{(Q)}$ in involution, with five independent integrals.

\begin{Remark}[rotationally invariant case]
The extension (\ref{KdV-deltaLn-phii-2}) gives a similar result, with
\begin{gather}\label{DWW-B1-t2-ht-rot}
h^{(Q)} = \frac{1}{2} \sum_{i=1}^N P_i^2-\alpha Q_1-\frac{1}{32} \left(16 Q_1^4+12 Q_1^2\sum_{i=2}^N Q_i^2+ \left(\sum_{i=2}^N Q_i^2\right)^2\right) +\frac{\beta}{{\bf Q}^2}.
\end{gather}
The integrals $h_{ij}$ are just replaced by the rotation algebra on the space $Q_2, \dots , Q_N$.
\end{Remark}

\subsubsection[Bi-Hamiltonian formulation when N=2]{Bi-Hamiltonian formulation when $\boldsymbol{N=2}$}\label{sec:DWW-N=2biHam}

Here we have
\begin{gather}\label{DWW-t2-f2-N=2}
 h^{(Q)} = \frac{1}{2} \big(P_1^2+P_2^2\big)-\alpha Q_1 -\frac{1}{32} \big(16 Q_1^4+12 Q_1^2 Q_2^2+Q_2^4\big) +\frac{\beta}{Q_2^2}.
\end{gather}

The Hamiltonian formulations (\ref{DWW-B0-t2-h}) and (\ref{DWW-t2-f2-N=2}) give coordinates ${\bf q}=(q_i,p_i,\alpha_0,\alpha_1)$ and ${\bf Q}=(Q_i,P_i,\beta,\alpha)$, respectively.
From the definitions of the canonical coordinates, we have
\begin{gather*}
 q_1=-\frac{1}{8} \big(6 Q_1^2+Q_2^2\big) ,\qquad q_2=Q_1,\qquad p_1=-\frac{1}{4} P_1, \nonumber\\
 p_2=\frac{1}{16} (Q_2 P_2-4Q_1 P_1), \qquad \alpha_0=- \frac{1}{8} h^{(Q)},\qquad \alpha_1=\frac{1}{16} \alpha.
\end{gather*}
Following the same procedure as Section \ref{sec:biHam-KdV-t1}, we introduce a degenerate extension of the canonical Poisson tensor in the 6D space with coordinates $\bf q$ and then use
$\big(\frac{\pa {\bf Q}}{\pa {\bf q}}\big) {\mathcal P}_0^{(q)} \big(\frac{\pa {\bf Q}}{\pa {\bf q}}\big)^{\rm T}= 16 {\mathcal P}_0^{(Q)}$, giving
\begin{subequations}
 \begin{gather}
{\mathcal P}_0^{(q)} = \left(
 \begin{matrix}
\phantom{-}0 & \phantom{-}0 & 1 & 0 & 0 & 0 \\
 \phantom{-}0 & \phantom{-}0 & 0 & 1 & 0 & 0 \\
 -1 & \phantom{-}0 & 0 & 0 & 0 & 0\\
 \phantom{-}0 & -1 & 0 & 0 & 0 & 0\\
 \phantom{-}0 & \phantom{-}0 & 0 & 0 & 0 & 0\\
 \phantom{-}0 & \phantom{-}0 & 0 & 0 & 0 & 0
 \end{matrix}
 \right), \nonumber\\
{\mathcal P}_0^{(Q)} = \left(
 \begin{matrix}
\phantom{-}0 & \phantom{-}0 & \phantom{-}0 & \phantom{-}\frac{1}{Q_2} & a_{15} & 0 \\[2mm]
 \phantom{-}0 & \phantom{-}0 & \phantom{-}\frac{1}{Q_2} & -\frac{2 Q_1}{Q_2^2} & a_{25} & 0 \\[2mm]
 \phantom{-}0 & -\frac{1}{Q_2} & \phantom{-}0 & \phantom{-}\frac{P_2}{Q_2^2} & a_{35} & 0 \\[2mm]
 -\frac{1}{Q_2} & \phantom{-}\frac{2 Q_1}{Q_2^2} & - \frac{P_2}{Q_2^2} & \phantom{-}0 & a_{45} & 0 \\[2mm]
 -a_{15} & -a_{25} & -a_{35} & -a_{45} & 0 & 0\\[2mm]
 \phantom{-}0 &\phantom{-}0 & \phantom{-}0 & \phantom{-}0 & 0 & 0
 \end{matrix}
 \right),\label{P0qQ-DWWt2}
\end{gather}
where the column $(a_{15},a_{25},a_{35},a_{45},0,0)^{\rm T} = -{\mathcal P}_1^{(Q)}\nabla_Q f^{(Q)}$, where $P_1^{(Q)}$ is the compatible (degenerate) canonical Poisson tensor on the $\bf Q$ space, which in turn leads to $\big(\frac{\pa {\bf q}}{\pa {\bf Q}}\big) {\mathcal P}_1^{(Q)} \big(\frac{\pa {\bf q}}{\pa {\bf Q}}\big)^{\rm T}= \frac{1}{32} {\mathcal P}_1^{(q)}$:
 \begin{gather}
{\mathcal P}_1^{(Q)} = \left(
 \begin{matrix}
 \phantom{-}0 & \phantom{-}0 & 1 & 0 & 0 & 0 \\
 \phantom{-}0 & \phantom{-}0 & 0 & 1 & 0 & 0 \\
 -1 & \phantom{-}0 & 0 & 0 & 0 & 0\\
 \phantom{-}0 & -1 & 0 & 0 & 0 & 0\\
 \phantom{-}0 & \phantom{-}0 & 0 & 0 & 0 & 0\\
 \phantom{-}0 & \phantom{-}0 & 0 & 0 & 0 & 0
 \end{matrix}
 \right), \nonumber\\
{\mathcal P}_1^{(q)} = \left(
 \begin{matrix}
 \phantom{-}0 & \phantom{-}0 & 12 q_2 & \phantom{-}b_{14} & b_{15} & 0 \\[1mm]
 \phantom{-}0 & \phantom{-}0 & -8 & -8 q_2 & b_{25} & 0 \\[1mm]
 -12 q_2 & \phantom{-}8 & \phantom{-}0 & \phantom{-}8 p_1 & b_{35} & 0\\[1mm]
 -b_{14} & \phantom{-}8 q_2 & -8 p_1 & \phantom{-}0 & b_{45} & 0 \\[1mm]
 -b_{15} & -b_{25} & -b_{35} & - b_{45} & 0 & 0 \\[1mm]
\phantom{-}0 & \phantom{-}0 & \phantom{-}0 & \phantom{-}0 & 0 & 0
 \end{matrix}
 \right),\label{P1Qq-DWWt2}
\end{gather}
where $b_{14}=4q_1+ 15 q_2^2$ and the column $(b_{15},b_{25},b_{35},b_{45},0,0)^{\rm T} = -4 {\mathcal P}_0^{(q)}\nabla_q h^{(q)}$.
\end{subequations}

We use the independent fluxes from (\ref{DWW-F12m}) to build the {\em bi-Hamiltonian ladders}:
\begin{gather*}
F_{20} = 8\alpha_1 = \frac{1}{2} \alpha,\qquad F_{21} = 2 \alpha_0 = -\frac{1}{4} h^{(Q)},\nonumber\\
F_{22} = h^{(q)} = \frac{1}{16} f^{(Q)},\qquad F_{23} = f^{(q)} = \frac{\beta}{64},
\end{gather*}
where $f^{(q)}$ is given by (\ref{DWW-B0-t2-F2j}) and $f^{(Q)}$ by (\ref{DWW-B1-t2-F1j}) (with $N=2$). These satisfy
\[
{\mathcal P}_1^{(q)}\nabla_q \alpha_1= {\mathcal P}_0^{(q)}\nabla_q \alpha_1=0 \qquad\text{and} \qquad {\mathcal P}_1^{(Q)}\nabla_Q \alpha= {\mathcal P}_0^{(Q)}\nabla_Q \alpha=0,
\]
together with
\begin{gather*}
{\mathcal P}_0^{(q)}\nabla_q \alpha_0= 0, \qquad {\bf q}_{t_h}={\mathcal P}_1^{(q)}\nabla_q \left(-\frac{1}{4}\alpha_0\right)={\mathcal P}_0^{(q)}\nabla_q h^{(q)}, \nn\\
 {\bf q}_{t_f}= {\mathcal P}_1^{(q)}\nabla_q \left(-\frac{1}{8} h^{(q)}\right) = P_0^{(q)} f^{(q)}, \quad {\mathcal P}_1^{(q)}\nabla_q f^{(q)} = 0, \\
 {\mathcal P}_0^{(Q)}\nabla_Q h^{(Q)} = 0, \qquad
 {\bf Q}_{t_h}={\mathcal P}_1^{(Q)}\nabla_Q h^{(Q)}={\mathcal P}_0^{(Q)}\nabla_Q f^{(Q)}, \nn\\
 {\bf Q}_{t_f}= {\mathcal P}_1^{(Q)}\nabla_Q f^{(Q)} = {\mathcal P}_0^{(Q)}\nabla_Q (- \beta ), \qquad {\mathcal P}_1^{(Q)}\nabla_Q \beta = 0. 
\end{gather*}

\begin{Remark}
The first two components of the $t_h$ and $t_f$ flows just reproduce the first two flows of the DWW hierarchy, analogous to equations (\ref{KdVt2-qthtf}) and (\ref{KdVt2-Qthtf}) for the stationary $t_2$ flow of the KdV hierarchy.
\end{Remark}

\subsection{The Lax representation}\label{sec:sDWW-Lax}

Following the approach of Section \ref{sec:sKdV-Lax}, we obtain the Lax representations.

\subsubsection[The t\_1 flow]{The $\boldsymbol{t_1}$ flow}

In this case our matrices are
\begin{gather}\label{DWW-UV1}
U = \left(
 \begin{matrix}
 0 & 1 \\
 \lambda^2-\lambda u_1-u_0 & 0
 \end{matrix}
 \right),\qquad V^{(1)} = \left(
 \begin{matrix}
 -\frac{1}{2}u_{1x} & 2\lambda+ u_1 \\
 a_{21} & \frac{1}{2}u_{1x}
 \end{matrix}
 \right).
\end{gather}
where $a_{21} = \big(\lambda^2-\lambda u_1-u_0\big) (2\lambda+u_1)-\frac{1}{2} a_{12xx}$, given in terms of the component $a_{12}$ of the matrix $V^{(1)}$.

{\it Using the coordinates of the Hamiltonian \eqref{DWW-B0-t1-h}}, we define
\begin{gather*}
L^{(1)} = -2 V^{(1)} = \begin{pmatrix}
 p_1 & -2 (2\lambda + q_1) \\
 a_{21} & -p_1
 \end{pmatrix} \quad\Rightarrow\quad z^2-2 h^{(q)}+4 \alpha_0 \lambda +16 \alpha_1 \lambda^2-16 \lambda^4=0,
\end{gather*}
where $a_{21} = \alpha_0-2 \alpha_1 q_1+\frac{1}{2}q_1^3+ \lambda \big(4 \alpha_1-q_1^2\big)+2 \lambda^2 q_1-4 \lambda^3$. The Lax representation of the equation generated by $h^{(q)}$ is
\begin{gather*}
L^{(1)}_x = \big\{L^{(1)},h^{(q)}\big\} = \big[U,L^{(1)}\big], \qquad\text{with}\quad U = \left(
 \begin{matrix}
 0 & 1 \\
 \lambda^2-\lambda q_1+\frac{3}{4} q_1^2-\alpha_1 & 0
 \end{matrix}
 \right) .
\end{gather*}

{\it Using the coordinates of \eqref{DWW-B1-t1-ht}}, we have
\begin{subequations}
\begin{gather}
L^{(1)} = V^{(1)} = \left(
 \begin{matrix}
\displaystyle\sum_{1}^N Q_iP_i & 2\lambda -{\bf Q}^2 \\
 a_{21} & \displaystyle-\sum_{1}^N Q_iP_i
 \end{matrix}
 \right) \nonumber\\
 \quad{}\Rightarrow\quad z^2+ 2\Phi^{(Q)} -4 \lambda h^{(Q)}+8 \alpha \lambda^2 -4 \lambda^4=0,\label{DWW-L1-phi}
\end{gather}
where
\[a_{21} = \sum_{1}^N \left(P_i^2+\frac{2\beta_i}{Q_i^2}\right) +\frac{1}{2}\lambda \big({\bf Q}^4-8 \alpha\big)+\lambda^2 {\bf Q}^2 +2 \lambda^3,\qquad {\bf Q}^2=\sum_1^N Q_i^2,\quad {\bf Q}^4=\left({\bf Q}^2\right)^2,\]
 with $\Phi^{(Q)}$ as given in (\ref{DWW-B1-t1-F1j}). The Lax equations generated by $h^{(Q)}$ and $\Phi^{(Q)}$ are
\begin{gather}
L^{(1)}_x = \big\{L^{(1)},h^{(Q)}\big\} = \big[U,L^{(1)}\big],\nonumber\\ \big\{L^{(1)}, \Phi^{(Q)}\big\} =0, \quad \text{with}\quad U =
 \begin{pmatrix}
 0 & 1 \\
 \lambda^2+\lambda {\bf Q}^2 +\frac{3}{4} {\bf Q}^4 -2\alpha & 0
 \end{pmatrix}.\label{DWW-L1x-phi}
\end{gather}
\end{subequations}
In fact, $\big\{L^{(1)}, h_{ij}\big\} = 0$, for each of the functions $h_{ij}$ of (\ref{KdV-stat-t1-hij}).

\begin{Remark}[rotationally symmetric case]
Similar equations hold for the case (\ref{DWW-B1-t1-ht-rot}).
\end{Remark}

\subsubsection[The t\_2 flow]{The $\boldsymbol{t_2}$ flow}

In this case our matrices are $U$ (the same as (\ref{DWW-UV1})) and
\begin{gather*}
 V^{(2)} = \left(
 \begin{matrix}
 -\frac{1}{2} \lambda u_{1x}-\frac{1}{4} (2 u_{0x}+3 u_1 u_{1x}) & 2\lambda^2+ \lambda u_1 +u_0+\frac{3}{4} u_1^2\\
 a_{21} & \frac{1}{2} \lambda u_{1x}+\frac{1}{4} (2 u_{0x}+3 u_1 u_{1x})
 \end{matrix}
 \right),
\end{gather*}
where
\[
a_{21} = \left(2\lambda^2+\lambda u_1+u_0+\frac{3}{4} u_1^2\right)\big(\lambda^2- \lambda u_1 - u_0\big)- \frac{1}{2} a_{12xx}.
\]

{\it Using the coordinates of the Hamiltonian \eqref{DWW-B0-t2-h}}, we define
\begin{gather*}
L^{(2)} = \frac{1}{2} V^{(2)} = \left(
 \begin{matrix}
 \lambda p_1 + p_2-q_2 p_1 & \lambda^2 +\frac{1}{2} q_2 \lambda+\frac{1}{8} \big(4 q_1+3 q_2^2\big) \\
 a_{21} & q_2 p_1-p_2 - \lambda p_1
 \end{matrix}
 \right) ,
\end{gather*}
where
\begin{gather*}
a_{21} = \lambda^4-\frac{1}{2} q_2 \lambda^3-\frac{1}{8} \big(4 q_1+q_2^2\big) \lambda^2+\frac{1}{4}\big(q_2^3+2 q_1q_2-16 \alpha_1\big)\lambda -\alpha_0\\
\hphantom{a_{21} =}{} +2 \alpha_1 q_2+\frac{1}{4}q_1^2-\frac{5}{64} q_2^4-p_1^2.
\end{gather*}

The characteristic equation of $L^{(2)}$ is
\begin{gather*}
z^2 +\frac{1}{2} f^{(q)} + \frac{1}{2} h^{(q)} \lambda + \alpha_0 \lambda^2 +4 \alpha_1 \lambda^3- \lambda^6=0,
\end{gather*}
where $f^{(q)}$ is given by (\ref{DWW-B0-t2-F2j}).

The Lax representation of the equations generated by $h^{(q)}$ and $f^{(q)}$ are
\begin{gather*}
L^{(2)}_x = \big\{L^{(2)},h^{(q)} \big\} = \big[U,L^{(2)} \big] , \qquad L^{(2)}_{t_1} = \big\{L^{(2)},f^{(q)} \big\} = \big[L^{(1)},L^{(2)} \big] ,
\end{gather*}
where $U$ and $L^{(1)}=\frac{1}{2} V^{(1)}$ (see (\ref{DWW-UV1})) are given by
\begin{gather*}
U = \left(
 \begin{matrix}
 0 & 1 \\
 \lambda^2-q_2 \lambda-q_1 & 0
 \end{matrix}
 \right), \qquad
L^{(1)} = \left(
 \begin{matrix}
 p_1 & \lambda+\frac{1}{2} q_2 \\
 a_{21} & -p_1
 \end{matrix}
 \right),
\end{gather*}
with
\[
a_{21} = \lambda^3-\frac{1}{2} q_2 \lambda^2 -\frac{1}{2} \big(2 q_1+q_2^2 \big) \lambda +\frac{1}{8} \big(5 q_2^3+8 q_1 q_2-32 \alpha_1 \big).
\]

{\it Using the coordinates of \eqref{DWW-B1-t2-ht}}, we have
\begin{subequations}
\begin{gather}\label{DWW-L2-phi}
L^{(2)} = 8 V^{(2)} = \left(
 \begin{matrix}
 \displaystyle\sum_{2}^N Q_iP_i -4 \lambda P_1 & 16\lambda^2 +8 \lambda Q_1- \displaystyle\sum_{2}^N Q_i^2 \\
 a_{21} & 4 \lambda P_1 - \displaystyle\sum_{2}^N Q_iP_i
 \end{matrix}
 \right) ,
\end{gather}
where
\begin{gather*}
a_{21} = \sum_{2}^N \left(P_i^2+\frac{2\beta_i}{Q_i^2}\right) -\lambda \left(4 \alpha+Q_1\left(2 Q_1^2+\sum_{2}^N Q_i^2\right)\right)\\
\hphantom{a_{21} =}{}
+\lambda^2 \left(4 Q_1^2+\sum_{2}^N Q_i^2\right) -8 Q_1 \lambda^3+16 \lambda^4.
\end{gather*}
The characteristic equation of $L^{(2)}$ is
\begin{gather}\label{DWW-L2-char-Q}
z^2 +2\Phi^{(Q)} +8 f^{(Q)} \lambda -32 h^{(Q)} \lambda^2+ 64 \alpha \lambda^3- 256\lambda^6=0,
\end{gather}
where $\Phi^{(Q)}$ and $f^{(Q)}$ are given in (\ref{DWW-B1-t2-F1j}).

The Lax equations generated by $h^{(Q)}$, $f^{(Q)}$ and $h_{ij}$ (not just the {\em combination} $\Phi^{(Q)}$) are
\begin{gather}
L^{(2)}_x = \big\{L^{(2)},h^{(Q)}\big\} = \big[U,L^{(2)}\big] , \qquad L^{(2)}_{t_1} = \big\{L^{(2)},f^{(Q)}\big\} = \big[L^{(1)},L^{(2)}\big] , \nonumber\\ \big\{L^{(2)}, h_{ij}\big\} = 0,\label{DWW-L2x-phi}
\end{gather}
where $L^{(1)}= -2 V^{(1)}$ (see (\ref{DWW-UV1})), written in the coordinates of (\ref{DWW-B1-t2-ht}), and $U$ are given by
\begin{gather}\label{DWW-L1QU}
L^{(1)} = \left(
 \begin{matrix}
 P_1 & -4\lambda -2 Q_1 \\
 a_{21} & -P_1
 \end{matrix}
 \right), \qquad U = \left(
 \begin{matrix}
 0 & 1 \\
 \lambda^2-\lambda Q_1+\frac{1}{8} \left(6 Q_1^2+\sum_{2}^N Q_i^2\right) & 0
 \end{matrix}
 \right),
\end{gather}
with
\[ a_{21} = \alpha+\frac{1}{2} Q_1 \sum_{1}^N Q_i^2 -\frac{1}{2} \lambda \left(2 Q_1^2+\sum_{2}^N Q_i^2\right) +2 Q_1 \lambda^2 -4 \lambda^3.
\]
\end{subequations}

\begin{Remark}[rotationally symmetric case]
Similar equations hold for the case (\ref{DWW-B1-t2-ht-rot}).
\end{Remark}

\section{The DWW stationary flows in tri-Hamiltonian form}\label{sec:DWW-triH}

We now use the Miura maps of Section \ref{sec:DWW-Miura} to construct a {\em tri-Hamiltonian} formulation of the {\em stationary flows} corresponding to (\ref{DWW-triHam}) and (\ref{DWW-wtn-vtn}). In this way, Figure \ref{B2B1B0-fig} is extended to an array of 9 Poisson matrices ${\mathcal P}_k^{(m)}$, $k,m=0,1,2$. The 3 Poisson matrices ${\mathcal P}_k^{(k)}$ can be {\em directly constructed} in canonical form, giving us coordinates $(q_i,p_i)$, $(Q_i,P_i)$ and $\big(\bar Q_i,\bar P_i\big)$, for $k=0,1,2$ respectively. Written in terms of these coordinates, the Miura maps then allow us to build the remaining 6 Poisson matrices.

\begin{enumerate}\itemsep=0pt
 \item The case $B_0^u$, has already been considered in Section \ref{sec:DWW-stat}, giving the Lagrangian (\ref{DWW-Lagn+2}), which we now write as
 \begin{gather}\label{DWW-Lagn+2-u}
{\mathcal L}_{n+2}^u = H_{n+2}^u - \alpha_0 u_1-\alpha_1 \big(4 u_0+u_1^2\big).
\end{gather}
The (generalised) Legendre transformation defines canonical coordinates $(q_i,p_i)$, the Poisson matrix ${\mathcal P}_0^{(0)}\equiv {\mathcal P}_0^{(q)}$ and the Hamiltonian function $h^{(q)}$.
 \item Using $B_1^w$, we have
\begin{gather}
B_1^w \delta_w H_{n+1}^w=0 \quad \Rightarrow\quad \delta_w \left(H_{n+1}^w-\beta_0 w_0-\beta_1 w_1\right)=0 \nonumber\\ \qquad {} \Rightarrow\quad {\mathcal L}_{n+1}^w = H_{n+1}^w-\beta_0 w_0-\beta_1 w_1.\label{B1wdH=0}
\end{gather}
The (generalised) Legendre transformation defines canonical coordinates $(Q_i,P_i)$, the Poisson matrix ${\mathcal P}_1^{(1)}\equiv {\mathcal P}_1^{(Q)}$ and the Hamiltonian function $h^{(Q)}$.
 \item Using $B_2^v$, we have
\begin{gather}\label{B2vdH=0}
B_2^v \delta_v H_{n}^v=0 \quad\Rightarrow\quad \delta_v \left(H_n^v-\gamma_0 v_0-\gamma_1 v_1\right)=0 \quad\Rightarrow\quad {\mathcal L}_n^v = H_n^v-\gamma_0 v_0-\gamma_1 v_1.\!\!\!
\end{gather}
The (generalised) Legendre transformation defines canonical coordinates $\big(\bar Q_i,\bar P_i\big)$, the Poisson matrix ${\mathcal P}_2^{(2)}\equiv {\mathcal P}_2^{(\bar Q)}$ and the Hamiltonian function $h^{(\bar Q)}$.
\end{enumerate}

\subsection[The t\_1 flow]{The $\boldsymbol{t_1}$ flow}

We now derive the explicit formulae for the case $n=1$.

{\it The $u$ space, with \eqref{DWW-Lagn+2-u}}:
We previously derived this as (\ref{DWW-B0-t1-h}), giving
\begin{gather}\label{DWW-B0-t1-hu}
h^{(q)} = \frac{1}{2} p_1^2-\frac{1}{2} q_1^4+2\alpha_1 q_1^2-\alpha_0 q_1,
\end{gather}
where $u_0 = \alpha_1-\frac{3}{4} u_1^2$, $u_1=q_1$, $u_{1x}=p_1$.

{\it The $w$ space, with \eqref{B1wdH=0}}:
Using $-H_2^w$, we have
\begin{subequations}
\begin{gather}\label{DWW-t1-L2w}
{\mathcal L}_{2}^w = \frac{1}{2} w_1 \big(w_{0x} + w_0^2 \big) -\frac{1}{8} w_1^3-\beta_0 w_0-\beta_1 w_1.
\end{gather}
This is degenerate and reduces to $w_0=\frac{w_{1x}+2 \beta_0}{2 w_1}$. Removing an exact derivative and an overall numerical factor, we find
\begin{gather}\label{DWW-t1-L2w-1}
{\mathcal L}_{2}^w = \frac{(w_{1x}+2 \beta_0)^2}{2 w_1} +\frac{1}{2} w_1 \big(w_1^2+ 8\beta_1 \big),
\end{gather}
corresponding to the Hamiltonian
\begin{gather}\label{DWW-B1-t1-hw}
h^{(Q)} = \frac{1}{2} Q_1 P_1^2-2 \beta_0 P_1-\frac{1}{2} Q_1 \big(8 \beta_1+Q_1^2 \big).
\end{gather}
\end{subequations}

{\it The $v$ space, with \eqref{B2vdH=0}}:
Using $H_1^v$, we have
\begin{subequations}
\begin{gather}\label{DWW-t1-L1v}
{\mathcal L}_{1}^v = -v_{0x}-v_0^2+\frac{1}{4} \big(2 v_0v_1+v_{1x} \big)^2-\gamma_0 v_0-\gamma_1 v_1.
\end{gather}
This is degenerate and reduces to $v_0=\frac{\gamma_0-v_1v_{1x}}{2 (v_1^2-1)}$. Removing an exact derivative and an overall numerical factor, we find
\begin{gather}\label{DWW-t1-L1v-1}
{\mathcal L}_{1}^v = \frac{1}{2 (1-v_1^2)} \big(v_{1x}^2-2 \gamma_0 v_1v_{1x}-4 \gamma_1 v_1 \big(1-v_1^2 \big)+\gamma_0^2 \big),
\end{gather}
corresponding to the Hamiltonian
\begin{gather}\label{DWW-B2-t1-hv}
h^{(\bar Q)} = \frac{1}{2} \big(1-\bar Q_1^2 \big) \bar P_1^2+ \gamma_0 \bar Q_1 \bar P_1+ 2 \gamma_1 \bar Q_1.
\end{gather}
\end{subequations}

\subsubsection{The Miura maps in these coordinates}

We consider the two steps induced by ${\bf w}\mapsto {\bf u}$ and ${\bf v}\mapsto {\bf w}$. We again extend each space to include the parameters as dynamical variables, which we define as ${\bf q} = (q_1,p_1,\alpha_0,\alpha_1)$, ${\bf Q} = (Q_1,P_1,\beta_0,\beta_1)$, ${\bf \bar Q} = \big(\bar Q_1,\bar P_1,\gamma_0,\gamma_1\big)$.

{\it The relation of $\bf q$ to $\bf Q$}: $q_1=Q_1$, $p_1=Q_1P_1-2 \beta_0$. The constraints on $u_0$ and $w_0$, together with $u_0=-w_{0x}-w_0^2$, gives $\alpha_1=-2 \beta_1$. The formula $\alpha_0=u_{1xx}-2 u_1^3+4 \alpha_1 u_1$ gives $\alpha_0=h^{(Q)}$. In summary
\begin{gather*}
q_1=Q_1,\qquad p_1=Q_1P_1-2 \beta_0,\qquad \alpha_0 = h^{(Q)}, \qquad \alpha_1= -2 \beta_1 \qquad\Rightarrow\qquad h^{(q)}=2 \beta_0^2.
\end{gather*}

{\it The relation of $\bf Q$ to $\bf \bar Q$}: $Q_1= -\bar P_1$, $P_1= \bar Q_1 \bar P_1-\gamma_0$. The ``$x$ derivative'' of the first gives $\big\{Q_1,h^{(Q)}\big\}=\big\{{-}\bar P_1,h^{(\bar Q)}\big\}$, which implies $\beta_0=-\gamma_1$. The formula $\beta_1= \frac{4 \beta_0 P_1-Q_1(P_1^2+3 Q_1^2)+2 Q_{1xx}}{8 Q_1}$ gives $\beta_1=-\frac{1}{4} h^{(\bar Q)}+\frac{1}{8} \gamma_0^2$. In summary
\begin{gather*}
Q_1=-\bar P_1,\qquad P_1=\bar Q_1 \bar P_1-\gamma_0,\qquad \beta_0 = -\gamma_1, \qquad \beta_1=-\frac{1}{4} h^{(\bar Q)}+\frac{1}{8} \gamma_0^2 \\
\qquad\Rightarrow\quad h^{(Q)}=-2 \gamma_0 \gamma_1.
\end{gather*}

\subsubsection{The array of Poisson brackets}\label{sec:DWW-t1-PiqQQb}

We constructed the 3 canonical representations (\ref{DWW-B0-t1-hu}), (\ref{DWW-B1-t1-hw}) and (\ref{DWW-B2-t1-hv}), respectively on spaces $\bf q$, $\bf Q$ and $\bf\bar Q$. The canonical brackets are then extended to include the parameters as Casimirs:
\begin{gather*}
{\mathcal{P}}^{(q)}_0={\mathcal{P}}^{(Q)}_1={\mathcal{P}}^{(\bar{Q})}_2=
 \left(\begin{matrix}
 \phantom{-}0&1&0&0 \\ -1&0&0&0 \\ \phantom{-}0&0&0&0\\\phantom{-}0&0&0&0
 \end{matrix}\right).
\end{gather*}
Following the same procedure as Section \ref{sec:DWW-N=1biHam}, we construct the other 6 Poisson brackets. From~$P_0^{(q)}$, we obtain
\begin{align*}
{\mathcal{P}}^{(Q)}_0 &=\frac{1}{4\beta_0} \left(\begin{matrix}
 \phantom{-}0& \phantom{-}2P_1& a_{13} &0\\
 - 2 P_1&\phantom{-}0& a_{23} &0\\
 - a_{13} & -a_{23} &0&0\\
 \phantom{-}0&\phantom{-}0&0&0
 \end{matrix}\right), \\
 {\mathcal{P}}^{(\bar Q)}_0 &=\frac{1}{4 \gamma_1^2} \left(\begin{matrix}
 0& \gamma_0\bar{Q}_1\bar{P}_1-2\gamma_1\bar{Q}_1-\gamma_0^2 & -\gamma_0 b_{13} & \gamma_1 b_{13}\\
 \gamma_0^2 + 2\gamma_1\bar{Q}_1 - \gamma_0\bar{Q}_1\bar{P}_1 & 0 & -\gamma_0 b_{23} & \gamma_1 b_{23} \\
 \gamma_0 b_{13} & \gamma_0 b_{23} & 0 & 0\\
 -\gamma_1 b_{13} & -\gamma_1 b_{23} & 0 & 0
 \end{matrix}\right) ,
\end{align*}
where $(a_{13},a_{23},0,0)^{\rm T} = {\mathcal{P}}^{(Q)}_1 \nabla_Q h^{(Q)}$ and $(b_{13},b_{23},0,0)^{\rm T} = {\mathcal{P}}^{(\bar Q)}_2 \nabla_{\bar Q} h^{(\bar Q)}$.

From $P_1^{(Q)}$, we obtain
\begin{gather*}
{\mathcal{P}}^{(q)}_1 = \left(\begin{matrix}
 \phantom{-}0&\phantom{-}q_1& a_{13} &0\\
 -q_1&\phantom{-}0& a_{23} &0\\
 -a_{13} &-a_{23} &0&0\\
 \phantom{-}0&\phantom{-}0&0&0
 \end{matrix}\right), \qquad
 {\mathcal{P}}^{(\bar Q)}_1 = \frac{1}{2\gamma_1} \left(\begin{matrix}
 \phantom{-}0 & \bar{Q}_1\bar{P}_1-\gamma_0 & b_{13} & 0\\
 \gamma_0-\bar{Q}_1\bar{P}_1 & \phantom{-}0 & b_{23} & 0 \\
 -b_{13} & -b_{23} & 0 & 0\\
 \phantom{-}0 & \phantom{-}0 & 0 & 0
 \end{matrix}\right) ,
\end{gather*}
where $(a_{13},a_{23},0,0)^{\rm T} = {\mathcal P}_0^{(q)} \nabla_q h^{(q)}$ and $(b_{13},b_{23},0,0)^{\rm T} = -{\mathcal P}_2^{(\bar Q)} \nabla_{\bar Q} h^{(\bar Q)}$.

From $P_2^{(\bar Q)}$, we obtain
\begin{gather*}
{\mathcal{P}}^{(Q)}_2 = \left(\begin{matrix}
 \phantom{-}0 & -Q_1 & 0 & a_{14}\\[1mm]
 \phantom{-}Q_1 & \phantom{-}0 & 0 & a_{24} \\
 \phantom{-}0 & \phantom{-}0 & 0 & 0 \\
 -a_{14} & -a_{24} & 0 & 0
 \end{matrix}\right), \qquad
 {\mathcal{P}}^{(q)}_2 = \left(\begin{matrix}
 \phantom{-}0&-q_1^2&0& b_{14}\\[2mm]
 \phantom{-}q_1^2&\phantom{-}0&0& b_{24}\\[1mm]
 \phantom{-}0&\phantom{-}0&0&0\\
 -b_{14} &-b_{24} &0&0
 \end{matrix}\right) ,
\end{gather*}
where $(a_{14},a_{24},0,0)^{\rm T} = -\frac{1}{4} {\mathcal{P}}^{(Q)}_1 \nabla_Q h^{(Q)}$ and $(b_{14},b_{24},0,0)^{\rm T} = \frac{1}{2} {\mathcal{P}}^{(q)}_0 \nabla_{q} h^{(q)}$.

Each of these has 2 Casimirs and the $t_1 =t_h$ flow in each space has a tri-Hamiltonian representation.

In the $\bf q$ space, we have
\begin{gather*}
 {\bf q}_{t_h} = {\mathcal{P}}_2^{(q)} \nabla_q (2\alpha_1) = {\mathcal{P}}_1^{(q)} \nabla_q \alpha_0 = {\mathcal{P}}_0^{(q)} \nabla_q h^{(q)}, \\
 {\mathcal{P}}_2^{(q)} \nabla_q \alpha_0 = {\mathcal{P}}_2^{(q)} \nabla_q h^{(q)} = {\mathcal{P}}_1^{(q)} \nabla_q h^{(q)} = {\mathcal{P}}_1^{(q)} \nabla_q \alpha_1
 = {\mathcal{P}}_0^{(q)} \nabla_q \alpha_1 = {\mathcal{P}}_0^{(q)} \nabla_q \alpha_0 = 0.
\end{gather*}

In the $\bf Q$ space, we have
\begin{gather*}
 {\bf Q}_{t_h} = {\mathcal{P}}_2^{(Q)} \nabla_Q (-4\beta_1) = {\mathcal{P}}_1^{(Q)} \nabla_Q h^{(Q)} = {\mathcal{P}}_0^{(Q)} \nabla_Q \left(2\beta_0^2\right),\\
 {\mathcal{P}}_2^{(Q)} \nabla_Q h^{(Q)} = {\mathcal{P}}_2^{(Q)} \nabla_Q \beta_0 = {\mathcal{P}}_1^{(Q)} \nabla_Q \beta_0 = {\mathcal{P}}_1^{(Q)} \nabla_Q \beta_1= {\mathcal{P}}_0^{(Q)} \nabla_Q \beta_1 = {\mathcal{P}}_0^{(Q)} \nabla_Q h^{(Q)} = 0.
\end{gather*}

In the $\bf \bar Q$ space, we have
\begin{gather*}
 {\bf \bar Q}_{t_h} = {\mathcal{P}}_2^{(\bar Q)} \nabla_{\bar Q} \left(h^{(\bar Q)}-\frac{1}{2} \gamma_0^2\right) = {\mathcal{P}}_1^{(\bar Q)} \nabla_{\bar Q} (-2 \gamma_0\gamma_1)
 = {\mathcal{P}}_0^{(\bar Q)} \nabla_{\bar Q} \left(2 \gamma_1^2\right),\\
 {\mathcal{P}}_2^{(\bar Q)} \nabla_{\bar Q} \gamma_0 = {\mathcal{P}}_2^{(\bar Q)} \nabla_{\bar Q} \gamma_1 = {\mathcal{P}}_1^{(\bar Q)} \nabla_{\bar Q} \gamma_1
 = {\mathcal{P}}_1^{(\bar Q)} \nabla_{\bar Q} \left(h^{(\bar Q)}-\frac{1}{2} \gamma_0^2\right) \\
\hphantom{{\mathcal{P}}_2^{(\bar Q)} \nabla_{\bar Q} \gamma_0}{}
 = {\mathcal{P}}_0^{(\bar Q)} \nabla_{\bar Q} \left(h^{(\bar Q)}-\frac{1}{2} \gamma_0^2\right) = {\mathcal{P}}_0^{(\bar Q)} \nabla_{\bar Q} (\gamma_0\gamma_1) = 0.
\end{gather*}

\begin{Remark}[relation of (\ref{DWW-B1-t1-hw}) to (\ref{DWW-f2-N=1})]
The canonical transformation
\begin{gather*}
Q_1 = \frac{1}{2} \tilde Q_1^2,\qquad P_1 = \frac{\tilde Q_1 \tilde P_1 + \sqrt{-2\beta}}{\tilde Q_1^2},\qquad\text{with} \quad \alpha= -4 \beta_1,\quad \beta = -8 \beta_0^2,
\end{gather*}
{\em gives} $h^{(Q)}= \frac{1}{2} h^{(\tilde Q)}$, where $h^{(\tilde Q)} = \frac{1}{2} \tilde P_1^2-\frac{1}{8} \tilde Q_1^6+\alpha \tilde Q_1^2+\frac{\beta}{\tilde Q_1^2}$, which is just (\ref{DWW-f2-N=1}) with relabelled variables.

This transformation is real when $\beta<0$.

Under this transformation, ${\mathcal P}_i^{(Q)}$ are transformed to ${\mathcal P}_i^{(\tilde Q)}$, with ${\mathcal P}_1^{(\tilde Q)}={\mathcal P}_1^{(Q)}$ and
\begin{gather*}
{\mathcal{P}}^{(\tilde Q)}_0 =
 2 \left(\begin{matrix}
 \phantom{-}0 & \phantom{-}\frac{1}{\tilde Q_1^2} & a_{13} & 0 \\
 -\frac{1}{\tilde Q_1^2} & \phantom{-}0 & a_{23} & 0 \\
 -a_{13} & -a_{23} & 0 & 0\\
 \phantom{-}0 & \phantom{-}0 & 0 & 0
 \end{matrix}
 \right), \qquad
 {\mathcal{P}}^{(\tilde Q)}_2 =
 \frac{1}{2} \left(\begin{matrix}
 \phantom{-}0 &- \tilde Q_1^2 & 0 & b_{14}\\[2mm]
 \phantom{-}\tilde Q_1^2 & \phantom{-}0 & 0 & b_{24}\\[1mm]
 \phantom{-}0&\phantom{-}0&0&0\\
 -b_{14} & -b_{24} &0&0
 \end{matrix}\right) ,
\end{gather*}
where $(a_{13},a_{23},0,0)^{\rm T} = -{\mathcal{P}}^{(\tilde Q)}_1 \nabla_{\tilde Q} h^{(\tilde Q)}$ and $(b_{14},b_{24},0,0)^{\rm T} = {\mathcal{P}}^{(\tilde Q)}_1 \nabla_{\tilde Q} h^{(\tilde Q)}$,
thus rendering the flow of $h^{(\tilde Q)}$ as tri-Hamiltonian
\begin{gather*}
\tilde {\bf Q}_{t_h} = {\mathcal{P}}_2^{(\tilde{Q})} \nabla_{\tilde Q} (2\alpha) = {\mathcal{P}}_1^{(\tilde{Q})} \nabla_{\tilde Q} h^{(\tilde{Q})}
 = {\mathcal{P}}_0^{(\tilde{Q})} \nabla_{\tilde Q} \left(-\frac{1}{2}\beta\right).
\end{gather*}
This system also has the Lax matrix (\ref{DWW-L1-phi}), with $N=1$.
\end{Remark}

\subsection[The t\_2 flow]{The $\boldsymbol{t_2}$ flow}

We now derive the explicit formulae for the case $n=2$.

{\it The $u$ space, with \eqref{DWW-Lagn+2-u}}:
We previously derived this as (\ref{DWW-B0-t2-h}), giving
\begin{subequations}
\begin{gather}\label{DWW-B0-t2-hu}
h^{(q)} = 5 q_2 p_1^2-4 p_1 p_2-\frac{q_2}{64} \big(4 q_1+q_2^2\big)\big(12 q_1+7 q_2^2\big)+\alpha_0 q_2 +\alpha_1\big(4 q_1+q_2^2\big),
\end{gather}
where $u_0=q_1$, $u_1=q_2$, $u_{0x} =2 (5q_2p_1-2 p_2)$, $u_{1x}=-4 p_1$. In~(\ref{DWW-B0-t2-F2j}), we also gave the first integral
\begin{gather}\label{DWW-B0-t2-fw}
f^{(q)} = \frac{1}{256} \big(4 q_1+3 q_2^2\big) \big(64(\alpha_0-2 \alpha_1 q_2)+64 p_1^2-16 q_1^2+5 q_2^4\big)-2 (p_2-q_2 p_1)^2.
\end{gather}
\end{subequations}

{\it The $w$ space, with \eqref{B1wdH=0}}:
Using $2H_3^w$, we have
\begin{subequations}
\begin{gather}\label{DWW-t2-L3w}
{\mathcal L}_{2}^w = \frac12w_{0x}^2-\frac18w_{1x}^2-\frac34w_1^2w_{0x}+\frac12w_0^4-\frac34w_0^2w_1^2+\frac5{32}w_1^4 -\beta_0 w_0-\beta_1 w_1.
\end{gather}
Defining $Q_1=w_0$, $Q_2=w_1$, $P_1=w_{0x}-\frac34w_1^2$, $P_2=-\frac14w_{1x}$, we obtain the Hamiltonian
\begin{gather}\label{DWW-B1-t2-hw}
h^{(Q)} = \frac{1}{2} P_1^2-2 P_2^2+\frac{3}{4} Q_2^2P_1-\frac{1}{2} Q_1^4+\frac{3}{4} Q_1^2Q_2^2+\frac{1}{8} Q_2^4+\beta_0 Q_1+\beta_1 Q_2.
\end{gather}
\end{subequations}

\begin{Remark}[first integral]
Rather than calculating fluxes of the modified PDEs, we derive the first integral $f^{(Q)}$ (also $f^{(\bar Q)}$ below) directly through the Miura maps in the stationary coordinates, later in our calculations.
\end{Remark}

{\it The $\boldsymbol{v}$ space, with \eqref{B2vdH=0}}:
Using $2 H_2^v$, we have
\begin{subequations}
\begin{gather}\label{DWW-t2-L2v}
{\mathcal L}_{2}^v = -\frac14v_{1x}^3-\frac32v_0v_1v_{1x}^2+v_{0x}v_{1x}-3v_0^2v_1^2v_{1x}+2v_0^3v_1\big(1-v_1^2\big)-\gamma_0 v_0-\gamma_1 v_1.
\end{gather}
Defining $\bar{Q}_1=v_0$, $\bar{Q}_2=v_1$, $\bar{P}_1=v_{1x}$, $\bar{P}_2=-\frac34v_{1x}^2-3v_0v_1v_{1x}+v_{0x}-3v_0^2v_1^2$, we obtain the Hamiltonian
\begin{gather}\label{DWW-B2-t2-hv}
h^{(\bar Q)} = \frac14\bar{P}_1^3+\bar{P}_1\bar{P}_2+\frac32\bar{Q}_1\bar{Q}_2\bar{P}_1^2+3\bar{Q}_1^2\bar{Q}_2^2\bar{P}_1 +2\bar{Q}_1^3\bar{Q}_2\big(\bar{Q}_2^2-1\big)+\gamma_0\bar{Q}_1+\gamma_1\bar{Q}_2.
\end{gather}
\end{subequations}

\subsubsection{The Miura maps in these coordinates}

We consider the two steps induced by ${\bf w}\mapsto {\bf u}$ and ${\bf v}\mapsto {\bf w}$. We again extend each space to include the parameters as dynamical variables, which we define as ${\bf q} = (q_1,q_2,p_1,p_2,\alpha_0,\alpha_1)$, ${\bf Q} = (Q_1,Q_2,P_1,P_2,\beta_0,\beta_1)$, ${\bf \bar Q} = \big(\bar Q_1,\bar Q_2,\bar P_1,\bar P_2,\gamma_0,\gamma_1\big)$.

{\it The relation of $\bf q$ to $\bf Q$}: found by using the Miura map and the definitions of $\bf q$ and $\bf Q$. Formulae for $\alpha_0$, $\alpha_1$ are derived from the equations of motion implied by $h^{(q)}$. In summary
\begin{gather*}
 q_1= -P_1-Q_1^2-\frac{3}{4} Q_2^2,\qquad\! q_2=Q_2,\qquad\! p_1=P_2,\qquad\! p_2= \frac{1}{2} Q_1 \big(P_1+Q_1^2\big)+ Q_2 P_2-\frac{1}{4} \beta_0, \\
 \alpha_1=\frac{1}{4} \beta_1,\qquad \alpha_0=\frac{1}{2} h^{(Q)}, \qquad h^{(q)}= -\frac{1}{2} f^{(Q)},\quad f^{(q)} = -\frac{1}{8} \beta_0^2,
\end{gather*}
where
\begin{gather*}
f^{(Q)} = \frac{1}{4} \big(P_1+Q_1^2\big) \big(8\beta_1+16 Q_1P_2+4 Q_2 P_1 +\big(8 Q_1^2+Q_2^2\big) Q_2\big) - \beta_0 (2 P_2+Q_1 Q_2),
\end{gather*}
satisfying $\big\{h^{(Q)},f^{(Q)}\big\}=0$.

{\it The relation of $\bf Q$ to $\bf \bar Q$}: found by using the Miura map and the definitions of~$\bf Q$ and~$\bf \bar Q$. Formulae for $\beta_0$, $\beta_1$ are derived from the equations of motion implied by $h^{(Q)}$. In summary
\begin{gather*}
 Q_1=\!\bar Q_1,\qquad Q_2=-\bar P_1- 2 \bar Q_1 \bar Q_2,\\ P_1=\!\bar P_2,\qquad P_2=\frac{1}{2} \big(\bar Q_1 \bar P_1+\bar Q_2 \bar P_2\big)+\frac{3}{2}\bar Q_1^2\bar Q_2-\frac{1}{4} \gamma_0, \\
 \beta_0= \gamma_1,\qquad \beta_1=\frac{1}{2} h^{(\bar Q)}, \qquad h^{(Q)}= \frac{1}{4} f^{(\bar Q)}-\frac{1}{8} \gamma_0^2,\qquad f^{(Q)} = \frac{1}{2} \gamma_0 \gamma_1,
\end{gather*}
where
\begin{gather*}
f^{(\bar Q)}=\big(\bar P_2+\bar Q_1^2\big)\big(\bar P_1\big(\bar P_1+ 4\bar Q_1 \bar Q_2\big) -2 \big(\bar P_2 -\bar Q_1^2\big)\big(\bar Q_2^2-1\big)\big)
 + 2 \gamma_0 \bar Q_2 \big(\bar P_2+\bar Q_1^2\big)\\ \hphantom{f^{(\bar Q)}=}-2 \gamma_1 \big(\bar Q_2 \bar P_1+2\bar Q_1 \big(\bar Q_2^2-1\big)\big),
\end{gather*}
satisfying $\big\{h^{(\bar Q)},f^{(\bar Q)}\big\}=0$.

\subsubsection{The array of Poisson brackets}\label{sec:DWW-t2-PiqQQb}

We constructed the 3 canonical representations (\ref{DWW-B0-t2-hu}), (\ref{DWW-B1-t2-hw}) and (\ref{DWW-B2-t2-hv}), respectively on spaces $\bf q$, $\bf Q$ and $\bf\bar Q$. The canonical brackets are then extended to include the parameters $\alpha_i$, $\beta_i$, $\gamma_i$ as Casimirs:
 \begin{gather*}
{\mathcal{P}}^{(q)}_0={\mathcal{P}}^{(Q)}_1={\mathcal{P}}^{(\bar{Q})}_2=
 \left(\begin{matrix}
 \phantom{-}0& \phantom{-}0&1&0&0&0 \\ \phantom{-}0& \phantom{-}0&0&1&0&0 \\ -1& \phantom{-}0&0&0&0&0 \\ \phantom{-}0&-1&0&0&0&0 \\ \phantom{-}0& \phantom{-}0&0&0&0&0\\ \phantom{-}0& \phantom{-}0&0&0&0&0
 \end{matrix}\right).
\end{gather*}
Following the same procedure as Section \ref{sec:DWW-t1-PiqQQb}, we construct the other 6 Poisson brackets.

From $P_0^{(q)}$, we obtain
\begin{gather*}
{\mathcal{P}}^{(Q)}_0 = \frac{1}{4\beta_0} \!\left(\!\begin{matrix}
 \phantom{-}0& -16Q_1 & 8(2 P_2+Q_1Q_2) & 4P_1\!+\!8 Q_1^2\!+\!3 Q_2^2 & a_{15} & 0\\
 16Q_1 & \phantom{-}0 & -32 Q_1^2 & \phantom{-}0 & a_{25} & 0\\
 -8(2 P_2+Q_1Q_2) & \phantom{-}32 Q_1^2 & \phantom{-}0 & \phantom{-}a_{34} & a_{35} & 0\\
 -\big(4P_1\!+\!8 Q_1^2\!+\!3 Q_2^2\big) & \phantom{-}0 & -a_{34} & \phantom{-}0 & a_{45} & 0 \\
 -a_{15} & - a_{25} & - a_{35} & -a_{45} & 0 & 0 \\
 \phantom{-}0 & \phantom{-}0 & \phantom{-}0 & \phantom{-}0 & 0 & 0
 \end{matrix}\!\right), \\
 {\mathcal{P}}^{(\bar Q)}_0 = \frac1{\gamma_1^2}\left(\!\begin{matrix}
 \phantom{-}0 & \phantom{-}b_{12} &\phantom{-}b_{13} &\gamma_1\big(2\bar{Q}_2\big(\bar{P}_2+\bar{Q}_1^2\big)-\gamma_0\big) &b_{15} &b_{16}\\
 -b_{12} &\phantom{-}0 &\phantom{-}b_{23} &\phantom{-}b_{24} &b_{25} &b_{26}\\
 -b_{13} & -b_{23} &\phantom{-}0 &\phantom{-}b_{34} &b_{35} &b_{36} \\
 \gamma_1\big(\gamma_0-2\bar{Q}_2\big(\bar{P}_2+\bar{Q}_1^2\big)\big) &-b_{24} & -b_{34} &\phantom{-}0 &b_{45} &b_{46} \\
 -b_{15} &-b_{25} &-b_{35} &-b_{45} &0 &0\\
 -b_{16} &-b_{26} &-b_{36} &-b_{46} &0 &0
 \end{matrix}\!\right),
\end{gather*}
where $a_{34} = -2 \big(2 \beta_0 +Q_1 \big(4 P_1+8 Q_1^2+3 Q_2^2\big)\big)$, $(a_{15},a_{25},a_{35},a_{45},0,0)^{\rm T} = 8 {\mathcal P}_1^{(Q)} \nabla_Q f^{(Q)}$ and
\begin{gather*}
 b_{12} = \gamma_0 \big(\bar P_1+2 \bar Q_1\bar Q_2\big)-2 \gamma_1 \bar Q_2^2,\qquad b_{13}=2\bar{Q}_1\big(2\gamma_1\big(\bar{Q}_2^2+1\big)-\gamma_0\big({\bar P}_1+2{\bar Q}_1{\bar Q}_2\big)\big),\nn\\
 b_{23}=2\big(\gamma_0({\bar P}_1+2{\bar Q}_1{\bar Q}_2\big)\bar{Q}_2-2\gamma_1\bar{Q}_2\big(\bar{Q}_2^2-1\big)-2\gamma_0\bar{Q}_1\big), \nn\\
 b_{24}=2\gamma_0\big({\bar P}_1+{\bar Q}_1{\bar Q}_2\big)\bar{Q}_1-2\gamma_1\big({\bar P}_1+4{\bar Q}_1{\bar Q}_2\big)\bar{Q}_2-2\gamma_0\bar{Q}_2\bar{P}_2+4\gamma_1\bar{Q}_1+\gamma_0^2 , \nn\\
 b_{34}=2\big(2\gamma_1\bar{Q}_1\bar{Q}_2\big({\bar P}_1+3{\bar Q}_1{\bar Q}_2\big)-2\gamma_0\big({\bar P}_1+{\bar Q}_1{\bar Q}_2\big)\bar{Q}_1^2 +\big(\gamma_0\bar{Q}_1-\gamma_1\bar{Q}_2\big)\big(2\bar{Q}_2\bar{P}_2-\gamma_0\big)\big), \nn\\
 (b_{15},b_{25},b_{35},b_{45},0,0)^{\rm T}={\mathcal{P}}^{(\bar{Q})}_2\nabla_{\bar{Q}}\big(\gamma_0 f^{({\bar Q})}-4\gamma_1 h^{({\bar Q})}\big),\nn\\
 (b_{16},b_{26},b_{36},b_{46},0,0)^{\rm T}=-{\mathcal{P}}^{(\bar{Q})}_2\nabla_{\bar{Q}}\big(\gamma_1 f^{({\bar Q})}\big). \nn
\end{gather*}

From $P_1^{(Q)}$, we obtain $\big(\frac{\pa\bf{q}}{\pa\bf{Q}}\big) {\mathcal{P}}^{(Q)}_1 \big(\frac{\pa\bf{q}}{\pa\bf{Q}}\big)^{\rm T} = -\frac{1}{8}{\mathcal{P}}^{(q)}_1$, of (\ref{P1Qq-DWWt2}), and
\begin{gather*}
 {\mathcal{P}}^{(\bar Q)}_1 = \frac1{\gamma_1} \left(\begin{matrix}
 \phantom{-}0&-\big(\bar P_1+2\bar Q_1 \bar Q_2\big)&2\bar{Q}_1\big(\bar P_1+2\bar Q_1 \bar Q_2\big)& \phantom{-}\gamma_1 &a_{15}&0\\
 \bar P_1+2\bar Q_1 \bar Q_2 & \phantom{-}0 & \phantom{-}a_{23} & \phantom{-}a_{24} &a_{25}&0\\
 -2\bar{Q}_1\big(\bar P_1+2\bar Q_1 \bar Q_2\big) & -a_{23} & \phantom{-}0 & \phantom{-}a_{34} &a_{35} &0 \\
 -\gamma_1 &-a_{24} & -a_{34} & \phantom{-}0 & a_{45} &0 \\
 -a_{15}&-a_{25}&-a_{35}&-a_{45}&0 &0\\
 \phantom{-}0& \phantom{-}0& \phantom{-}0& \phantom{-}0&0&0
 \end{matrix}\right),
\end{gather*}
where
\begin{gather*}
	a_{23} = 2\big(2\bar{Q}_1-\bar{Q}_2\bar P_1-2\bar Q_1 \bar Q_2^2\big), \qquad a_{24}=2\big(\bar{Q}_2\bar{P}_2-\bar{Q}_1\bar{P}_1-\bar{Q}_1^2\bar{Q}_2\big)-\gamma_0, \nn\\ a_{34}=2\big(2\bar{Q}_1^2\big({\bar P}_1+{\bar Q}_1{\bar Q}_2\big)-2\bar{Q}_1\bar{Q}_2\bar{P}_2+\gamma_0\bar{Q}_1-\gamma_1\bar{Q}_2\big),\nn\\
(a_{15},a_{25},a_{35},a_{45},0,0)^{\rm T}=-{\mathcal{P}}^{(\bar{Q})}_2\nabla_{\bar{Q}}f^{({\bar Q})}.
\end{gather*}

From $P_2^{(\bar Q)}$, we obtain $\big(\frac{\pa {\bf Q}}{\pa {\bf \bar Q}}\big) {\mathcal P}_2^{(\bar Q)} \big(\frac{\pa {\bf Q}}{\pa {\bf \bar Q}}\big)^{\rm T}= \frac{1}{2} {\mathcal P}_2^{(Q)}$ and
$\big(\frac{\pa {\bf q}}{\pa {\bf Q}}\big) {\mathcal P}_2^{(Q)} \big(\frac{\pa {\bf q}}{\pa {\bf Q}}\big)^{\rm T}= \frac{1}{4} {\mathcal P}_2^{(q)}$:
\begin{gather*}
{\mathcal{P}}^{(Q)}_2 = \left(\begin{matrix}
 \phantom{-}0 & -2 & \phantom{-}0 & \phantom{-}Q_1 & 0 & a_{16}\\[1mm]
 \phantom{-}2 & \phantom{-}0 & -4 Q_1 & - Q_2 & 0 & a_{26} \\
 \phantom{-}0 & \phantom{-}4 Q_1 & \phantom{-}0 & -(P_1+3 Q_1^2) & 0 & a_{36} \\
 - Q_1 & \phantom{-}Q_2 & P_1+3 Q_1^2 & \phantom{-}0 & 0 & a_{46} \\
 \phantom{-}0 & \phantom{-}0 & \phantom{-}0 & \phantom{-}0 & 0 & 0\\
 -a_{16} & -a_{26} & -a_{36} & -a_{46} & 0 & 0
 \end{matrix}\right),\\
 {\mathcal{P}}^{(q)}_2 = \left(\begin{matrix}
 \phantom{-}0 & \phantom{-}0 & 3 q_2^2-4 q_1 & \frac{1}{2} q_2 (4 q_1+15 q_2^2) & b_{15} & b_{16}\\
 \phantom{-}0 & \phantom{-}0 & -4 q_2 & -4 q_1-7 q_2^2 & b_{25} & b_{26}\\
 4 q_1-3 q_2^2 & \phantom{-}4 q_2 & \phantom{-}0 & \phantom{-}4 q_2 p_1 & b_{35} & b_{36} \\
 -\frac{1}{2} q_2 (4 q_1+15 q_2^2) & 4 q_1+7 q_2^2 & -4 q_2 p_1 & \phantom{-}0 & b_{45} & b_{46} \\
 -b_{15} & -b_{25} & - b_{35} & -b_{45} & 0 & 0 \\
 -b_{16} & -b_{26} & - b_{36} & -b_{46} & 0 & 0
 \end{matrix}\right),
\end{gather*}
where
\begin{gather*}(a_{16},\dots ,0)^{\rm T} = {\mathcal P}_1^{(Q)} \nabla_Q h^{(Q)},\qquad (b_{15},\dots ,0)^{\rm T} = 4 {\mathcal P}_0^{(q)} \nabla_q f^{(q)}, \\ (b_{16},\dots ,0)^{\rm T} = {\mathcal P}_0^{(q)} \nabla_q h^{(q)}.
\end{gather*}

Each of these has 2 Casimirs and the flows in each space have tri-Hamiltonian representations.

In the $\bf q$ space, we have
\begin{gather*}
{\bf q}_{t_h} = {\mathcal{P}}_2^{(q)} \nabla_q \alpha_1 = {\mathcal{P}}_1^{(q)} \nabla_q \left(-\frac{1}{4}\alpha_0\right) = {\mathcal{P}}_0^{(q)} \nabla_q h^{(q)}, \\ 
 {\bf q}_{t_f} = {\mathcal{P}}_2^{(q)} \nabla_q \left(\frac{1}{4}\alpha_0\right) = {\mathcal{P}}_1^{(q)} \nabla_q \left(-\frac{1}{8} h^{(q)}\right) = {\mathcal{P}}_0^{(q)} \nabla_q f^{(q)},\\ 
{\mathcal{P}}_2^{(q)} \nabla_q h^{(q)} = {\mathcal{P}}_2^{(q)} \nabla_q f^{(q)} = {\mathcal{P}}_1^{(q)} \nabla_q f^{(q)} = {\mathcal{P}}_1^{(q)} \nabla_q \alpha_1
 = {\mathcal{P}}_0^{(q)} \nabla_q \alpha_1 = {\mathcal{P}}_0^{(q)} \nabla_q \alpha_0 = 0. 
\end{gather*}

In the $\bf Q$ space, we have
\begin{gather*}
{\bf Q}_{t_h} = {\mathcal{P}}_2^{(Q)} \nabla_Q \beta_1 = {\mathcal{P}}_1^{(Q)} \nabla_Q h^{(Q)} = {\mathcal{P}}_0^{(Q)} \nabla_Q \left(-\frac{1}{2} f^{(Q)}\right), \\ 
{\bf Q}_{t_f} = {\mathcal{P}}_2^{(Q)} \nabla_Q (- h^{(Q)}) = {\mathcal{P}}_1^{(Q)} \nabla_Q f^{(Q)} = {\mathcal{P}}_0^{(Q)} \nabla_Q \left(\frac{1}{4} \beta_0^2\right), \\ 
 {\mathcal{P}}_2^{(Q)} \nabla_Q f^{(Q)} = {\mathcal{P}}_2^{(Q)} \nabla_Q \beta_0 = {\mathcal{P}}_1^{(Q)} \nabla_Q \beta_0 = {\mathcal{P}}_1^{(Q)} \nabla_Q \beta_1
 = {\mathcal{P}}_0^{(Q)} \nabla_Q \beta_1 = {\mathcal{P}}_0^{(Q)} \nabla_{Q} h^{(Q)} = 0 . 
\end{gather*}

In the $\bf \bar Q$ space, we have
\begin{gather*}
	 {\bf \bar Q}_{t_h} = {\mathcal{P}}_2^{(\bar Q)} \nabla_{\bar Q} h^{(\bar Q)} = {\mathcal{P}}_1^{(\bar Q)} \nabla_{\bar Q} \left(\frac{1}{4} f^{(\bar Q)}-\frac{1}{8} \gamma_0^2\right)
 = {\mathcal{P}}_0^{(\bar Q)} \nabla_{\bar Q} \left(-\frac{1}{4} \gamma_0 \gamma_1\right), \\ 
 {\bf \bar Q}_{t_f} = {\mathcal{P}}_2^{(\bar Q)} \nabla_{\bar Q} f^{(\bar Q)} = {\mathcal{P}}_1^{(\bar Q)} \nabla_{\bar Q} \left(- \gamma_0 \gamma_1\right)
 = {\mathcal{P}}_0^{(\bar Q)} \nabla_{\bar Q} \left(-\frac{1}{2} \gamma_1^2\right), \\ 
 {\mathcal{P}}_2^{(\bar Q)} \nabla_{\bar Q} \gamma_0 = {\mathcal{P}}_2^{(\bar Q)} \nabla_{\bar Q} \gamma_1 = {\mathcal{P}}_1^{(\bar Q)} \nabla_{\bar Q} \gamma_1 = {\mathcal{P}}_1^{(\bar Q)} \nabla_{\bar Q} h^{(\bar Q)} \nn\\
\hphantom{{\mathcal{P}}_2^{(\bar Q)} \nabla_{\bar Q} \gamma_0}{}
 = {\mathcal{P}}_0^{(\bar Q)} \nabla_{\bar Q} h^{(\bar Q)} = {\mathcal{P}}_0^{(\bar Q)} \nabla_{\bar Q} \left(f^{(\bar Q)}-\frac{1}{2} \gamma_0^2\right) = 0 . 
\end{gather*}

\begin{Remark}[relation of (\ref{DWW-B1-t2-hw}) to (\ref{DWW-t2-f2-N=2})]\label{Rem:3HamQuartic}
The canonical transformation
\begin{subequations}
\begin{gather}
Q_1 = \frac{2^\frac{2}{3}\big(\sqrt{-2\beta}-\tilde{Q}_2\tilde{P}_2\big)}{\tilde{Q}_2^2},\qquad Q_2=2^\frac{2}{3}\tilde{Q}_1,\nn\\
 P_1 = \frac{\tilde Q_2^2}{2^\frac53}-2^{\frac{4}{3}} \frac{\big(\sqrt{-2\beta}-\tilde Q_2 \tilde P_2\big)^2}{\tilde Q_2^4},\qquad P_2=\!\frac{\tilde{P}_1}{2^{\frac{2}{3}} },\label{DWW-f2-N=2-cantran}
\end{gather}
with $\alpha= \beta_1$, $\beta = -\frac{1}{2} \beta_0^2$, {\em gives} $h^{(Q)}= -2^\frac{2}{3} h^{(\tilde Q)}$, $f^{(Q)}=-2^\frac{1}{3} f^{(\tilde{Q})}$, where
\begin{gather}
 h^{(\tilde Q)} =\frac{1}{2} \big(\tilde{P}_1^2+\tilde{P}_2^2\big)-\alpha \tilde{Q}_1
 -\frac{1}{32}\big(16\tilde{Q}_1^4+12\tilde{Q}_1^2\tilde{Q}_2^2+\tilde{Q}_2^4\big)+\frac{\beta}{\tilde{Q}_2^2}, \label{DWW-f2-N=2tilde} \\
 f^{(\tilde{Q})} = \tilde{P}_1(\tilde{Q}_1\tilde{P}_1+\tilde{Q}_2\tilde{P}_2)-2\tilde{Q}_1h^{(\tilde{Q})}-\frac{\alpha}2\big(4\tilde{Q}_1^2 +\tilde{Q}_2^2\big)\nn\\
 \hphantom{ f^{(\tilde{Q})} =}{}-\frac1{16}\tilde{Q}_1\big(4\tilde{Q}_1^2+\tilde{Q}_2^2\big)\big(4\tilde{Q}_1^2+3\tilde{Q}_2^2\big), \label{DWW-f2-N=2ftilde}
\end{gather}
which are just (\ref{DWW-t2-f2-N=2}) and (\ref{DWW-B1-t2-F1j}) (for $N=2$), with relabelled variables $\tilde {\bf Q} = \big(\tilde Q_i,\tilde P_i,\beta,\alpha\big)$.

This transformation is real when $\beta <0$.

Under this transformation, ${\mathcal P}_i^{(Q)}$ are transformed to ${\mathcal P}_i^{(\tilde Q)}$, with ${\mathcal P}_1^{(\tilde Q)}={\mathcal P}_1^{(Q)}$ and (up to overall numerical factors)
\begin{gather}
{\mathcal{P}}^{(\tilde Q)}_0=
 \begin{pmatrix}
 \phantom{-}0& \phantom{-}0& \phantom{-}0& \phantom{-}\frac1{\tilde{Q}_2}& a_{15} & 0\\[2mm]
 \phantom{-}0& \phantom{-}0& \phantom{-}\frac1{\tilde{Q}_2}&-\frac{2\tilde{Q}_1}{\tilde{Q}_2^2}& a_{25} & 0\\[2mm]
 \phantom{-}0&-\frac1{\tilde{Q}_2}& \phantom{-}0& \phantom{-}\frac{\tilde{P}_2}{\tilde{Q}_2^2} &a_{35} & 0\\[2mm]
 -\frac1{\tilde{Q}_2}& \phantom{-}\frac{2\tilde{Q}_1}{\tilde{Q}_2^2}&-\frac{\tilde{P}_2}{\tilde{Q}_2^2}& \phantom{-}0& a_{45} & 0\\[2mm]
 -a_{15}&-a_{25}&-a_{35}&-a_{45}&0&0 \\[2mm]
 \phantom{-} 0& \phantom{-}0& \phantom{-}0& \phantom{-}0&0&0
 \end{pmatrix}, \nonumber\\
 {\mathcal{P}}^{(\tilde Q)}_2 =
 \begin{pmatrix}
 \phantom{-}0& \phantom{-}0& \phantom{-}2\tilde{Q}_1& \phantom{-}\tilde{Q}_2&0 &b_{16}\\[2mm]
 \phantom{-}0& \phantom{-}0& \phantom{-}\tilde{Q}_2& \phantom{-}0&0 &b_{26}\\[2mm]
 -2\tilde{Q}_1&-\tilde{Q}_2& \phantom{-}0& \phantom{-}\tilde{P}_2&0 &b_{36}\\[2mm]
 -\tilde{Q}_2& \phantom{-}0&-\tilde{P}_2& \phantom{-}0&0 &b_{46}\\[2mm]
 \phantom{-}0& \phantom{-}0& \phantom{-}0& \phantom{-}0&0&0\\[2mm]
 -b_{16}&-b_{26}&-b_{36}&-b_{46}&0&0
 \end{pmatrix},\label{DWW-t2-P02Qtilde}
\end{gather}
where $(a_{15},a_{25},a_{35},a_{45},0,0)^{\rm T}= -{\mathcal{P}}_1^{(\tilde{Q})} \nabla_{\tilde Q}f^{(\tilde{Q})}$ and
$(b_{16},b_{26},b_{36},b_{46},0,0)^{\rm T}=2{\mathcal{P}}_1^{(\tilde{Q})}\nabla_{\tilde Q}h^{(\tilde{Q})}$,
thus rendering the flow of (\ref{DWW-f2-N=2tilde}) as tri-Hamiltonian:
\begin{gather}\label{DWW-t2-Qtilde-lad}
\tilde {\bf Q}_{t_h} = {\mathcal{P}}_2^{(\tilde{Q})} \nabla_{\tilde Q} \left(\frac{1}{2}\alpha\right) = {\mathcal{P}}_1^{(\tilde{Q})} \nabla_{\tilde Q} h^{(\tilde{Q})}
 = {\mathcal{P}}_0^{(\tilde{Q})} \nabla_{\tilde Q} f^{(\tilde{Q})}.
\end{gather}
This system also has the Lax matrix (\ref{DWW-L2-phi}), with $N=2$.
\end{subequations}
\end{Remark}

\section{Generalisations: coupling with the Calogero--Moser model}\label{sec:generalisations}

Several of our Lax pairs can be generalised (when $N=3$) to incorporate an arbitrary function in the Hamiltonian, which allows us to relate the system to the rational Calogero--Moser model, following an approach described in \cite{f22-1}.

In Section \ref{sec:generalisations-QP}, we present generalisations of the Garnier and H\'enon--Heiles systems, as well as the Hamiltonian (\ref{DWW-B1-t2-ht}), with quartic potential, but others, such as (\ref{DWW-B1-t1-ht}), can be similarly generalised. For each case we give three functions, $h^{(B)}$, $f^{(B)}$ and $h_{23}^{(B)}$, which are in involution, along with Lax matrices.

The connection to the Calogero--Moser model is explained in Section \ref{sec:generalisations-CM}.

\subsection[Generalisations in (Q\_i,P\_i) coordinates]{Generalisations in $\boldsymbol{(Q_i,P_i)}$ coordinates}\label{sec:generalisations-QP}

\subsubsection{Generalised Garnier system}

In Section \ref{sec:KdV-stat-t1}, we derived two versions of the multicomponent Garnier system, given by (\ref{L1-t1-phi}) and (\ref{L1-t1-phi-2}). For brevity, we just give the generalisation of the second of these:
\begin{gather*}
 h^{(B)} =\frac{1}{2} \big(P_1^2+P_2^2+P_3^2\big)+ k\big(Q_1^2+Q_2^2+Q_3^2\big)^2+\frac{\beta}{Q_1^2+Q_2^2+Q_3^2} + \frac{1}{Q_2^2}B\left(\frac{Q_3}{Q_2}\right), \\ 
f^{(B)} = (Q_1P_2-Q_2P_1)^2+(Q_2P_3-Q_3P_2)^2+(Q_3P_1-Q_1P_3)^2+2\frac{Q_1^2
+Q_2^2+Q_3^2}{Q_2^2}B\!\left(\frac{Q_3}{Q_2}\right), \\ 
h_{23}^{(B)} =(Q_2 P_3-Q_3 P_2)^2 +2 \frac{\big(Q_2^2+Q_3^2\big)}{Q_2^2} B\left(\frac{Q_3}{Q_2}\right), 
\end{gather*}
where $B\big(\frac{Q_3}{Q_2}\big)$ is an arbitrary function. When $k=-\frac{1}{4}$ and $B\big(\frac{Q_3}{Q_2}\big)=0$, these reduce to (\ref{L1-t1-phi-2}) and elements of the rotation algebra.

The Lax matrix (\ref{L1-phi}), with $N=3$ and $\sum\frac{2\beta_i}{Q_i^2}$ replaced by $\frac{2\beta}{{\bf Q}^2}$, can similarly be generalised:
\begin{gather*}
L^{(1)} =\begin{pmatrix}
 Q_1 P_1+Q_2 P_2+Q_3 P_3 & 2\lambda -Q_1^2-Q_2^2-Q_3^2 \\
 a_{21} & -(Q_1 P_1+Q_2 P_2+Q_3 P_3) \\
 \end{pmatrix} \nonumber\\
 \qquad \Rightarrow\quad z^2+16k\lambda^3-4 h^{(B)}\lambda+ f^{(B)} +2\beta=0,
\end{gather*}
where
\[
a_{21}= -4k\lambda \big(2\lambda+Q_1^2+Q_2^2+Q_3^2\big) +P_1^2+P_2^2+P_3^2 +\frac{2\beta}{Q_1^2+Q_2^2+Q_3^2}+\frac{2}{Q_2^2}B\left(\frac{Q_3}{Q_2}\right).
\]

The Lax equations (\ref{L1x-phi}) are unchanged, but $U$ is deformed:
\begin{gather*}
 U= \left(\begin{matrix}
 0&1\\
 -4k\big(\lambda+Q_1^2+Q_2^2+Q_3^2\big)&0
 \end{matrix} \right) \quad\Rightarrow\quad \left\{ \begin{matrix}
 L^{(1)}_x = \big\{L^{(1)},h^{(B)}\big\} = \big[U,L^{(1)}\big] , \\[1mm]
 \big\{L^{(1)}, f^{(B)}\big\} = \big\{L^{(1)}, h_{23}^{(B)}\big\}= 0.
 \end{matrix} \right.
\end{gather*}

\subsubsection{Generalised H\'enon--Heiles system}\label{sec:generalisations-QP-HH}

The generalised H\'enon--Heiles potential of (\ref{L2-ht}) was considered in \cite[Section~3.5]{f22-1}, for the case $N=3$, with some additional terms:
\begin{subequations}\label{hfBU3}
\begin{gather}
 h^{(B)} =\frac{1}{2} \big(P_1^2\!+\!P_2^2\!+\!P_3^2\big)\!+\!\frac{\omega^2}{2}\big(4Q_1^2\!+\!Q_2^2\!+\!Q_3^2\big)\! +\! k Q_1 \big(2 Q_1^2\!+\!Q_2^2\!+\!Q_3^2\big)\! +\! \frac{1}{Q_2^2}B\!\left(\frac{Q_3}{Q_2}\right), \label{hBU3} \\
f^{(B)} =P_1 (Q_1P_1+Q_2P_2+Q_3P_3) - 2 Q_1 h^{(B)} +2\omega^2 Q_1\big(2Q_1^2+Q_2^2+Q_3^2\big) \nn \\
\hphantom{f^{(B)} =}{}
 + \frac{k}{4} \big(16 Q_1^4+ 12 Q_1^2 \big(Q_2^2+Q_3^2\big) + \big(Q_2^2+Q_3^2\big)^2\big), \label{fBU3} \\
h_{23}^{(B)} =(Q_2 P_3-Q_3 P_2)^2 +2 \frac{\big(Q_2^2+Q_3^2\big)}{Q_2^2} B\left(\frac{Q_3}{Q_2}\right), \label{h23BU3}
\end{gather}
\end{subequations}
where $B\left(\frac{Q_3}{Q_2}\right)$ is an arbitrary function.

When $B(z) = \beta_2+\frac{\beta_3}{z^2}$, $k=\frac{1}{2}$, $\omega=0$, then $h^{(B)}$ reduces to $h^{(Q)}$ of (\ref{L2-ht}). We also have that~$f^{(B)}$ reduces to $f^{(Q)}$ and $h_{23}^{(B)}$ to $h_{23}+2(\beta_2+\beta_3)$.

The Lax matrix (\ref{L2-Lax-ht}), with $N=3$, can similarly be generalised:
\begin{subequations}
\begin{gather}\label{L2-Lax-ht-B}
L^{(2)} = \left(
 \begin{matrix}
 Q_2 P_2+Q_3 P_3 - 4 \lambda P_1 & 16\lambda^2 + 8 \lambda Q_1-Q_2^2-Q_3^2 \\
 a_{21} & 4 \lambda P_1 -Q_2 P_2-Q_3 P_3
 \end{matrix}
 \right) ,
\end{gather}
where
\begin{gather*}
a_{21}= 2k \big(16 \lambda^3- 8 Q_1 \lambda^2\!+ \lambda \big(4 Q_1^2+Q_2^2+Q_3^2\big)\big)+8\omega^2 \lambda (Q_1-2\lambda)+P_2^2+P_3^2+\frac{2}{Q_2^2} B\!\left(\frac{Q_3}{Q_2}\right)\!,
\end{gather*}
 and the characteristic equation is
\begin{gather}\label{L2-Lax-ht-char-B}
z^2= 512 k \lambda^5 - 256 \omega^2 \lambda^4 +32 h^{(B)} \lambda^2-8 f^{(B)} \lambda -h_{23}^{(B)}.
\end{gather}
The Lax equations are just the same as (\ref{L2-laxeq-Q}):
\begin{gather}
L^{(2)}_x = \big\{L^{(2)},h^{(B)}\big\} = \big[U,L^{(2)}\big],\qquad L^{(2)}_{t_1} = \big\{L^{(2)},f^{(B)}\big\} = \big[L^{(1)},L^{(2)}\big], \nn\\ \big\{L^{(2)},h_{23}^{(B)}\big\}=0,\label{L2-laxeq-B}
\end{gather}
but with deformed versions of $U$ and $L^{(1)}$:
\begin{gather}\label{L2-U-L1-B}
U= \left(\begin{matrix}
 0&1\\
 2k(\lambda-Q_1)-\omega^2&0
 \end{matrix}\right),\qquad
L^{(1)}= \left(\begin{matrix}
 P_1 & -2(2\lambda+Q_1)\\
 b_{21} & -P_1
 \end{matrix}\right),
\end{gather}
where
\[
b_{21} = k\big({-}8\lambda^2+4\lambda Q_1 -\big(2Q_1^2+Q_2^2+Q_3^2\big)\big)+2\omega^2(2\lambda-Q_1).
\]
\end{subequations}

\subsubsection{Generalisation of the Hamiltonian (\ref{DWW-B1-t2-ht})}

The potential of (\ref{DWW-B1-t2-ht}) was labelled $U_4$ in \cite[Section 3.3]{f22-1}, for the case $N=3$. With some additional terms, and writing $\beta=\frac{1}{2} \omega^2$, we have
\begin{subequations}
\begin{gather}
 h^{(B)} = \frac{1}{2} \big(P_1^2+P_2^2+P_3^2\big)-\alpha Q_1 +\frac{\omega^2}{2}\big(4Q_1^2+Q_2^2+Q_3^2\big) \nn\\
\hphantom{h^{(B)} =}{}
+ k \big(16Q_1^4+12Q_1^2\big(Q_2^2+Q_3^2\big)+\big(Q_2^2+Q_3^2\big)^2\big) + \frac{1}{Q_2^2}B\left(\frac{Q_3}{Q_2}\right), \label{hBU4} \\
f^{(B)} = P_1 (Q_1P_1+Q_2P_2+Q_3P_3) - 2 Q_1 h^{(B)}-\frac{\alpha}{2}\big(4Q_1^2+Q_2^2+Q_3^2\big) \nn \\
\hphantom{f^{(B)} =}{}
 +2\omega^2 Q_1\big(2Q_1^2+Q_2^2+Q_3^2\big) +2k Q_1\big(4Q_1^2+3Q_2^2+3Q_3^2\big)\big(4Q_1^2+Q_2^2+Q_3^2\big), \label{fBU4} \\
h_{23}^{(B)} = (Q_2 P_3-Q_3 P_2)^2 +2 \frac{\big(Q_2^2+Q_3^2\big)}{Q_2^2} B\left(\frac{Q_3}{Q_2}\right), \label{h23BU4}
\end{gather}
\end{subequations}
where $B\big(\frac{Q_3}{Q_2}\big)$ is an arbitrary function.

When $B(z) = \beta_2+\frac{\beta_3}{z^2}$, $k=-\frac{1}{32}$, $\omega=0$, then $h^{(B)}$ reduces to $h^{(Q)}$ of (\ref{DWW-B1-t2-ht}). We also have that $f^{(B)}$ reduces to $f^{(Q)}$ and $h_{23}^{(B)}$ to $h_{23}+2(\beta_2+\beta_3)$.

The Lax matrix (\ref{DWW-L2-phi}), with $N=3$, is similarly generalised:
\begin{gather*}
L^{(2)} = \left(
 \begin{matrix}
 Q_2 P_2 + Q_3 P_3 -4 \lambda P_1 & 16\lambda^2 +8 \lambda Q_1- \big(Q_2^2 + Q_3^2\big) \\
 a_{21} & 4 \lambda P_1 -(Q_2 P_2 + Q_3 P_3)
 \end{matrix}
 \right) ,
\end{gather*}
where
\begin{gather*}
a_{21} = P_2^2 + P_3^2+\frac{2}{Q_2^2}B\left(\frac{Q_3}{Q_2}\right)+4 \lambda \big(2 \omega^2 (Q_1-2 \lambda)- \alpha\big) \nn\\
\hphantom{a_{21} =}{}
 -32 k \lambda \big(16 \lambda^3-8 \lambda^2 Q_1 +\lambda \big(4 Q_1^2+Q_2^2 + Q_3^2\big)- Q_1 \big(2 Q_1^2+Q_2^2 + Q_3^2\big)\big). \nn
\end{gather*}
The characteristic equation of $L^{(2)}$ is
\begin{gather*}
z^2+8192k\lambda^6+256 \omega^2\lambda^4+64\alpha\lambda^3-32 h^{(B)}\lambda^2+8 f^{(B)} \lambda+ h_{23}^{(B)}=0.
\end{gather*}
The Lax equations generated by $h^{(B)}$, $f^{(B)}$ and $h_{23}^{(B)}$ are the same as (\ref{DWW-L2x-phi}):
\begin{gather*}
L^{(2)}_x =\big \{L^{(2)},h^{(B)}\big\} = \big[U,L^{(2)}\big] , \qquad L^{(2)}_{t_1} = \big\{L^{(2)},f^{(B)}\big\} = \big[L^{(1)},L^{(2)}\big] , \nn\\ \big\{L^{(2)}, h_{23}^{(B)}\big\} = 0,
\end{gather*}
where $L^{(1)}$ and $U$ are deformations of (\ref{DWW-L1QU}):
\begin{gather*}\label{DWW-L1QU-B}
L^{(1)} = \begin{pmatrix}
 P_1 & -4\lambda -2 Q_1 \\
 b_{21} & -P_1 \\
 \end{pmatrix}
 , \qquad U = \begin{pmatrix}
 0 & 1 \\
 -32 k \big(\lambda^2-\lambda Q_1+\frac{1}{8} \big(6 Q_1^2+Q_2^2+Q_3^2\big)\big)-\omega^2 & 0 \\
 \end{pmatrix},
\end{gather*}
with
\begin{gather*}
b_{21} = \alpha +2\omega^2 (2\lambda-Q_1)\\
\hphantom{b_{21} =}{}
 -32 k \big(\tfrac{1}{2} Q_1 \big(Q_1^2+Q_2^2+Q_3^2\big) -\tfrac{1}{2} \lambda \big(2 Q_1^2+Q_2^2+Q_3^2\big) +2 Q_1 \lambda^2 -4 \lambda^3\big).
\end{gather*}

\begin{Remark}
The cases (\ref{hBU3}) and (\ref{hBU4}) belong to a family of Hamiltonians which are separable in {\em generalised parabolic coordinates} \cite{f22-1}.
\end{Remark}

\subsection{Transformation to Calogero--Moser coordinates}\label{sec:generalisations-CM}

The Hamiltonian $h^{(B)}$, with just the $B$ term (for example, (\ref{hBU3}) with $k=\omega=0$) is known \cite{08-8, 07-10} to have 4 integrals for general $B$, with an additional integral for the particular choice $B(z)=\frac{9g^2(1+z^2)^2}{2(1-3z^2)^2}$. This particular choice, with the (orthogonal) canonical transformation, generated by
\begin{gather*}
S=\frac{1}{\sqrt{3}}(q_1+q_2+q_3)P_1+\frac{1}{\sqrt{2}}(q_1-q_2)P_2+\frac{1}{\sqrt{6}}(q_1+q_2-2q_3)P_3,
\end{gather*}
gives the rational Calogero--Moser potential:
\begin{gather*}
\frac{1}{Q_2^2}B\left(\frac{Q_3}{Q_2}\right) = g^2\left(\frac{1}{(q_1-q_2)^2}+\frac{1}{(q_1-q_3)^2}+\frac{1}{(q_2-q_3)^2}\right).
\end{gather*}
In \cite{f22-1}, we gave a large class of additional potentials which could be added to $B$ and thus coupled with the rational Calogero--Moser system, but had no Lax representations. In this paper, we see that any system written in Section~\ref{sec:generalisations-QP} gives rise to a~{\em coupling} of the rational Calogero--Moser system {\em with a Lax representation}. To illustrate this, we just present the H\'enon--Heiles case.

\subsubsection{Calogero--Moser system, coupled with the H\'enon--Heiles potential}\label{CM-with-HH}

Since the formulae are more complicated in the Calogero--Moser coordinates, we introduce three functions which give some simplifications:
\begin{gather*}
\tau=q_1+q_2+q_3,\qquad \rho=q_1^2+q_2^2+q_3^2, \qquad \delta=(q_1-q_2)^2+(q_1-q_3)^2+(q_2-q_3)^2.
\end{gather*}
Under this canonical transformation, the functions $h^{(B)}$, $f^{(B)}$, $h_{23}^{(B)}$ (of (\ref{hfBU3})) take the form
\begin{gather*}
h^{{\rm (CM)}} = \frac{1}{2}\big(p_1^2 + p_2^2 + p_3^2\big) + g^2 \left(\frac{1}{(q_1 - q_2)^2} + \frac{1}{(q_1 - q_3)^2}
+ \frac{1}{(q_2 - q_3)^2}\right) \\
\hphantom{h^{{\rm (CM)}} =}{}
+ \frac{k}{3\sqrt{3}}\tau(6\rho - \delta) + \frac{\omega^2}2(4\rho - \delta) , \\ 
f^{{\rm (CM)}} = \frac{1}{\sqrt{3}}(p_1+p_2+p_3)(q_1p_1+q_2p_2+q_3p_3)-\frac{2\tau}{\sqrt{3}} h^{{\rm (CM)}} \nn\\
\hphantom{f^{{\rm (CM)}} =}{}
 +\frac{2}{3\sqrt{3}}\omega^2\tau(6\rho-\delta)+\frac{k}{36}\big(144\rho^2+5\delta^2-60\rho\delta\big),\\ 
h_{23}^{\rm (CM)} = \frac{1}{3} ((q_2 - q_3)p_1 + (q_3 - q_1)p_2 + (q_1 - q_2)p_3)^2 \\
\hphantom{h_{23}^{\rm (CM)} =}{}
+ \frac{2}{3}g^2\delta \left(\frac{1}{(q_1 - q_2)^2}
+ \frac{1}{(q_1 - q_3)^2} + \frac{1}{(q_2 - q_3)^2}\right) . 
\end{gather*}

The Lax matrices of Section \ref{sec:generalisations-QP-HH} take the form
\begin{gather*}
 L^{(2)}=\begin{pmatrix}
 a_{11}& a_{12}\\
 a_{21}&-a_{11}
 \end{pmatrix}, \qquad
 U= \begin{pmatrix}
 0 & 1\\
 2k\left(\lambda-\frac1{\sqrt{3}}\tau\right)-\omega^2 & 0
 \end{pmatrix}, \nonumber\\
 L^{(1)}=\frac1{\sqrt{3}}\begin{pmatrix}
 p_1+p_2+p_3 & -4\sqrt{3}\lambda-2\tau\\
 \sqrt{3}b_{21}&-(p_1+p_2+p_3)
 \end{pmatrix},
\end{gather*}
where
\begin{gather*}
 a_{11}= q_1p_1+q_2p_2+q_3p_3-\frac{1}{3}\big(\tau+4 \sqrt{3} \lambda\big) (p_1+p_2+p_3),\qquad a_{12}= 16\lambda^2+\frac{8}{\sqrt{3}}\tau\lambda-\frac{1}{3} \delta , \nn\\
 a_{21} = 32k \left(\lambda^3 -\frac{1}{2\sqrt{3}} \tau \lambda^2+\frac{1}{16} \lambda (4\rho-\delta)\right)+\frac{8\omega^2\lambda}{\sqrt{3}} \big(\tau-2\sqrt{3} \lambda\big) \nn\\
\hphantom{a_{21} =}{}
 +\frac{2}{3}\big(p_1^2+p_2^2+p_3^2-p_1p_2-p_1p_3-p_2p_3\big)\\
\hphantom{a_{21} =}{}
 +2g^2\left(\frac{1}{(q_1-q_2)^2}+\frac{1}{(q_1-q_3)^2}+\frac{1}{(q_2-q_3)^2}\right), \nn \\
 b_{21} = \frac{2\omega^2}{\sqrt{3}} \big(2 \sqrt{3} \lambda - \tau\big) -k \left(8\lambda^2-\frac{4\tau}{\sqrt{3}} \lambda +\frac{1}{3} (6\rho -\delta)\right). \nn
\end{gather*}
The characteristic equation of $L^{(2)}$ and the Lax equations are just the same as (\ref{L2-Lax-ht-char-B}) and (\ref{L2-laxeq-B}), but written in terms of these coordinates. Setting $k=0$, the system reduces to the resonant harmonic oscillator case presented in \cite[Section~4.2]{f22-1}.

This Lax matrix is certainly not as elegant as the usual Calogero--Moser one \cite{75-3}, but it {\em does} include the additional potentials. Furthermore, the Calogero--Moser system is known to be superintegrable \cite{83-11} and $f^{(CM)}$, $h_{23}^{(CM)}$ are related to his ``additional'' integrals, rather than those generated by the usual Lax matrix.

\section{Conclusions}

In this paper we have reconsidered the relationship of integrable nonlinear evolution equations and their stationary flows, which define finite-dimensional Hamiltonian systems (also integrable).

Multicomponent squared eigenfunction expansions gave us Hamiltonians with higher degrees of freedom, such as (\ref{L1-t1-phi-2}), (\ref{L2-ht}), (\ref{DWW-B1-t1-ht}), (\ref{DWW-B1-t2-ht}) and their rotationally symmetric versions. Restricting the dimension allowed us to build Poisson maps, giving us bi-Hamiltonian representations, by comparing the definitions of the corresponding canonical variables. The space is extended to include some arbitrary parameters as {\em dynamical variables} and the Poisson maps are {\em non-canonical}. For the DWW hierarchy, this was extended to a {\em tri-Hamiltonian} representation in Section \ref{sec:DWW-triH}. This used the DWW Miura maps, so (for each time-evolution) gave 3 tri-Hamiltonian systems. In particular, we showed that the well known Hamiltonian (\ref{DWW-f2-N=2tilde}), with quartic potential (see \cite{83-12,82-8}), is tri-Hamiltonian.

For each of these systems, we presented a $2 \times 2$ Lax representation, derived from the zero-curvature representation of the corresponding coupled KdV equation. For the case of 3 degrees of freedom, these Lax representations are further generalised in Section \ref{sec:generalisations} to connect with the results of~\cite{f22-1}, where we studied the class of Hamiltonians separable by {\em generalised parabolic coordinates}. A large subclass of these have the arbitrary function $B\big(\frac{Q_3}{Q_2}\big)$, introduced in Section~\ref{sec:generalisations}. This gave a~way of deriving Lax pairs and first integrals for a coupling of the rational Calogero--Moser model to many other potential functions, such as the H\'enon--Heiles potential, given in Section~\ref{CM-with-HH}.

From their derivation, the Lax matrices of this paper inherit a polynomial $\lambda$-dependence, with complicated coefficients. Clearly a larger matrix with linear $\lambda$-dependence (such as a~deformation of the usual Lax matrix of~\cite{75-3}) would be preferable. Currently these generalisations are restricted to 3 degrees of freedom, related to the results of \cite{f22-1}. For higher degrees of freedom, we would like generalisations with simple reductions to the 3 degrees of freedom case. These are quite varied and not necessarily related to the Calogero--Moser model.

In this paper we only considered the cases $M=1$ and $M=2$ in (\ref{e-Lax}), mainly presenting details of the first two nontrivial stationary flows. The squared eigenfunction representations~(\ref{KdV-deltaLn-phii-1}) and (\ref{DWW-dH=phi2}), corresponding to $h^{(Q)}$ of Sections \ref{sec:KdV-stat} and \ref{sec:DWW-stat}, is the most interesting.

The $t_1$ flow has the general structure
\begin{subequations}
\begin{gather}\label{t1-conc}
h^{(Q)} = \frac{1}{2} \sum_{i=1}^N \left(P_i^2+\frac{2\beta_i}{Q_i^2}\right) + U\left(\sum_{i=1}^N Q_i^2\right),
\end{gather}
and has a universal set of integrals, given by (\ref{KdV-stat-t1-hij}). For $N=3$, this is the first case in Table~2 of~\cite{90-22}. For general $N$, this class of Hamiltonian (on a curved space background) has been analysed in~\cite{09-10}, where the same set of universal integrals was derived. The particular form of the function $U$ in (\ref{t1-conc}), derived in our construction, depends upon the value of $M$ in (\ref{e-Lax}), but universally possesses a Lax representation.

The $t_2$ flow has the general structure
\begin{gather}\label{t2-conc}
h^{(Q)} =\frac{1}{2}P_1^2+ \frac{1}{2} \sum_{i=2}^N \left(P_i^2+\frac{2\beta_i}{Q_i^2}\right) + U\left(Q_1,\sum_{i=2}^N Q_i^2\right),
\end{gather}
\end{subequations}
which also has the universal set of integrals (\ref{KdV-stat-t1-hij}), but for $2\leq i<j\leq N$, as well as the integral~$f^{(Q)}$. For $N=2$ we just have the integral $f^{(Q)}$, but we see that the potential belongs to the class separable in parabolic coordinates (see \cite[equation~(2.2.41)]{90-16}). From our construction, we have two additional features: $M+1$ compatible Poisson brackets (for $N=2$) and a Lax representation (for all~$N$).

The $t_2$ flow is particularly interesting and leads to a number of important questions: Can we extend the multi-Poisson formulation beyond $N=2$? Even for $N=2$, can the multi-Poisson formulation be extended to the entire class of potentials separable in parabolic coordinates? Can the Lax pair be similarly extended? Some insight into these questions may be obtained by considering the case of (\ref{e-Lax}) with $M=3$.

\subsection*{Acknowledgements}

This work was supported by the National Natural Science Foundation of China (grant no. 11871396).

\pdfbookmark[1]{References}{ref}
\LastPageEnding

\end{document}